\documentclass[twocolumn,prd,noshowpacs,nofootinbib,amsmath,amssymb,superscriptaddress]{revtex4}
\usepackage{graphicx,epsfig,psfrag,bm,amssymb}
\usepackage{dcolumn}
\usepackage{bm}
\usepackage{color}
\usepackage{mathrsfs,amsmath, amsfonts,hepunits, color}
\usepackage{tikz}
\usepackage{slashed}
\usepackage{cancel}
\usepackage{mciteplus}
\usetikzlibrary{arrows,shapes}
\usetikzlibrary{trees}
\usetikzlibrary{matrix,arrows} 				

\usepackage{slashed}

\newcommand{\be}{\begin{eqnarray}}
\newcommand{\ee}{\end{eqnarray}}
\def\lsim{\mathrel{\rlap{\lower4pt\hbox{\hskip 0.5 pt$\sim$}}
    \raise1pt\hbox{$<$}}}                
\def\gsim{\mathrel{\rlap{\lower4pt\hbox{\hskip1pt$\sim$}}
    \raise1pt\hbox{$>$}}}

\def\lsim{\mathrel{\rlap{\lower4pt\hbox{\hskip1pt$\sim$}}
    \raise1pt\hbox{$<$}}}
\def\gsim{\mathrel{\rlap{\lower4pt\hbox{\hskip1pt$\sim$}}
    \raise1pt\hbox{$>$}}}

\newcommand{\apr}{A^{\prime}}

\begin{document}

\title{Discovering Inelastic Thermal-Relic Dark Matter at Colliders}
\author{Eder Izaguirre}
\author{ Gordan Krnjaic}
\author{ Brian Shuve}
\affiliation{Perimeter Institute for Theoretical Physics, Waterloo, Ontario, Canada    }

\begin{abstract}
Dark Matter particles with inelastic interactions are ubiquitous in extensions of the 
Standard Model, yet remain challenging to fully probe with existing strategies. We propose a series of powerful  searches at hadron and lepton colliders that are sensitive to inelastic dark matter dynamics. In representative models featuring either a massive dark photon or a magnetic dipole interaction, we find that the LHC and BaBar could offer strong sensitivity to the thermal-relic  dark matter
parameter space for dark matter masses between $\sim$ 100 MeV--100 GeV and fractional 
mass-splittings above the percent level; future searches at Belle II with a dedicated monophoton trigger could  also offer
 sensitivity to  thermal-relic scenarios with masses below a few GeV.
 Thermal scenarios with either larger masses or splittings are largely ruled out; lower masses remain viable yet may be accessible with other search strategies. 
\end{abstract}

\maketitle

\section{Introduction}
\label{sec:intro}

The observed cosmic abundance of dark matter (DM) \cite{Agashe:2014kda,Ade:2015xua}  is clear evidence of physics beyond the Standard Model (SM). While the  non-gravitational dynamics of DM are not currently known, additional interactions with the SM are well-motivated and arise in many theories. In thermal DM scenarios - a compelling paradigm for DM physics - these possible interactions  account for the observed cosmological abundance via DM annihilation into the SM.    This framework motivates DM physics with both mass scales and interaction strengths potentially accessible at current experiments. It is therefore imperative to robustly test the thermal DM paradigm with broad and complementary experimental approaches.

In the most commonly studied weakly interacting massive particle (WIMP) DM scenario, there is only one DM particle which interacts with the SM via a single type of interaction. In this case, obtaining the observed DM abundance through thermal freeze-out  fixes a minimum coupling between SM and DM states, and a variety of experiments can be used to test the possibility of thermal DM. These constraints are strongest for DM with masses above a few GeV, and a combination of direct-detection, indirect-detection, collider, beam-dump, astrophysical, and cosmological probes can dramatically narrow the window for thermal DM \cite{Cushman:2013zza,Essig:2013lka,Gershtein:2013iqa,Pierce:2014spa}. 

 However, the dark side of particle physics could exhibit a richer structure, especially given the complexity of the SM, and DM could even live in a dark sector (DS) with additional particles and forces \cite{Boehm:2003hm, Pospelov:2007mp, Pospelov:2008zw, ArkaniHamed:2008qn, Strassler:2006im}. 
This presents both new challenges and new opportunities:~some probes of DM can be dramatically less sensitive in even the simplest DS scenarios, relaxing the constraints on thermal DM, while at the same time new prospects for the discovery of DM  emerge.

In this paper, we explore some of the striking signatures at colliders that can  appear in a generic DS. A representative example of a DS consists of a dark matter particle which is charged under a hidden gauge or global symmetry. The DM can have both a symmetry-preserving  mass and, if the symmetry is spontaneously broken, also a symmetry-violating  mass, which  splits the mass eigenstates. In the limit that the symmetry-breaking mass is much smaller than the symmetry-preserving mass, the DM interactions are off-diagonal (between different mass eigenstates). This is a straightforward realization of the inelastic DM (iDM) scenario proposed by Tucker-Smith and Weiner \cite{TuckerSmith:2001hy}, with profound implications for experimental probes of DM. In particular, the abundance of the heavier eigenstate can be large in the early universe, facilitating efficient co-annihilation of DM, whereas the heavier eigenstate is depleted today, suppressing indirect- and direct-detection signatures. The small DM halo velocities imply that DM has insufficient energy to up-scatter into the heavier state, and so interactions through the off-diagonal coupling are ineffective today.

\begin{figure}[t!]
\vspace{0.5cm}
\includegraphics[width=8.8cm]{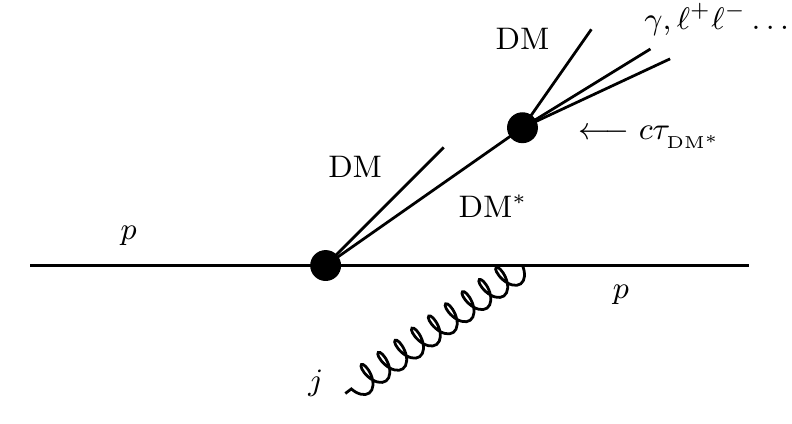} 
\caption{Schematic diagram depicting a characteristic iDM  production event at the LHC:~$p p \to j + {\rm DM} ~{\rm DM}^*$, where DM is the dominant
DM component in our halo and DM$^*$ is a heavier, unstable DS  state. The final state contains  visible SM particles ($\gamma$ or $\ell^+\ell^-)$ and missing transverse
energy produced in association with a QCD jet. At lepton colliders, a similar process of interest is $e^+e^- \to \gamma ~{\rm DM}   ~ {\rm DM^*} $.}
\label{fig:cartoonfigure}
\end{figure}

By contrast, the energies of colliders such as the LHC and $B$-factories are typically large enough to produce both the lighter and heavier DM mass eigenstates. In the iDM scenario, the dominant DM coupling to SM states is through the off-diagonal interaction, and so both eigenstates are produced simultaneously. When the heavier, ``excited'' dark state (denoted with an asterisk) decays to the lighter, ``ground'' dark state, some visible SM states are emitted. Thus, in addition to the standard DM missing transverse energy ($\cancel{E}_{\rm T}$) collider signature \cite{Petriello:2008pu,Gershtein:2008bf,Cao:2009uw,Beltran:2010ww, Bai:2010hh,Fox:2011fx,Fox:2011pm,Rajaraman:2011wf, Bai:2011jg,Goodman:2010ku,Goodman:2010yf,Bai:2012xg,Carpenter:2012rg,Lin:2013sca,Aad:2013oja,Carpenter:2013xra,Khachatryan:2014rwa,Aad:2014tda,Berlin:2014cfa, Aad:2014vka,ATLAS:2014wra,Khachatryan:2014tva,Izaguirre:2014vva,Khachatryan:2014rra,Aad:2014vea,Aad:2015zva,Haisch:2015ioa}, where the DM system recoils off of a jet, photon, vector boson, or Higgs boson, iDM models typically feature the emission of associated soft SM states \cite{Bai:2011jg,Weiner:2012cb}.
The characteristic  iDM collider signature is the production of $\mathrm{DM}+\mathrm{DM}^*$ in association with a hard SM object $X$, followed by the subsequent decay of $\mathrm{DM^*}\rightarrow\mathrm{DM} +Y$ for some potentially different SM states $Y$. The production is summarized as
\be \label{eq:schematic-production}
pp  &\to&   X +  \mathrm{DM}+\mathrm{DM}^*   \nonumber \\     &\to&  X + \mathrm{DM}+ \biggl( \mathrm{DM}^* \to \mathrm{DM}+  { Y}\biggr)     \equiv  X + \slashed{E}_{\rm T}+ Y ~,~ \nonumber
\ee
 and is depicted schematically in Fig. \ref{fig:cartoonfigure}. $X$ is any state that can be used to trigger on the event and reconstruct $\slashed{E}_{\rm T}$; throughout this study, we consider the case where $X$ is a jet for hadron colliders, and $X$ is a photon for lepton colliders. $Y$ depends on the mode by which DM couples inelastically to the SM. As we elaborate in Sec.~\ref{sec:models}, the  representative models we consider lead to two promising modes of $\mathrm{DM}^*$ decay, namely $Y=\gamma$ and $Y=\ell^+\ell^-$.
 
\begin{figure*}[t!] 
 \vspace{0.cm}
 \includegraphics[width=8.6cm]{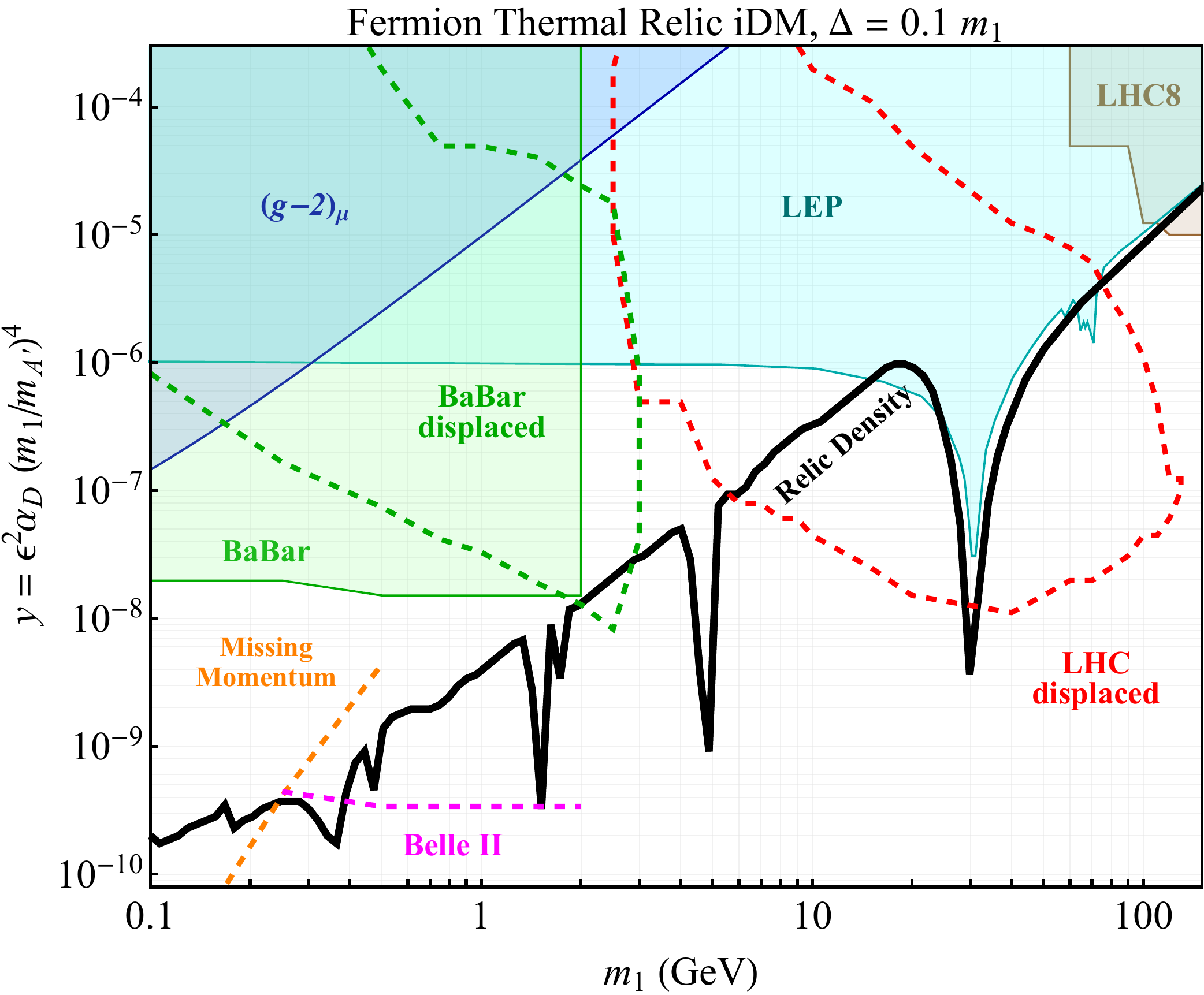}  
 \includegraphics[width=8.6 cm]{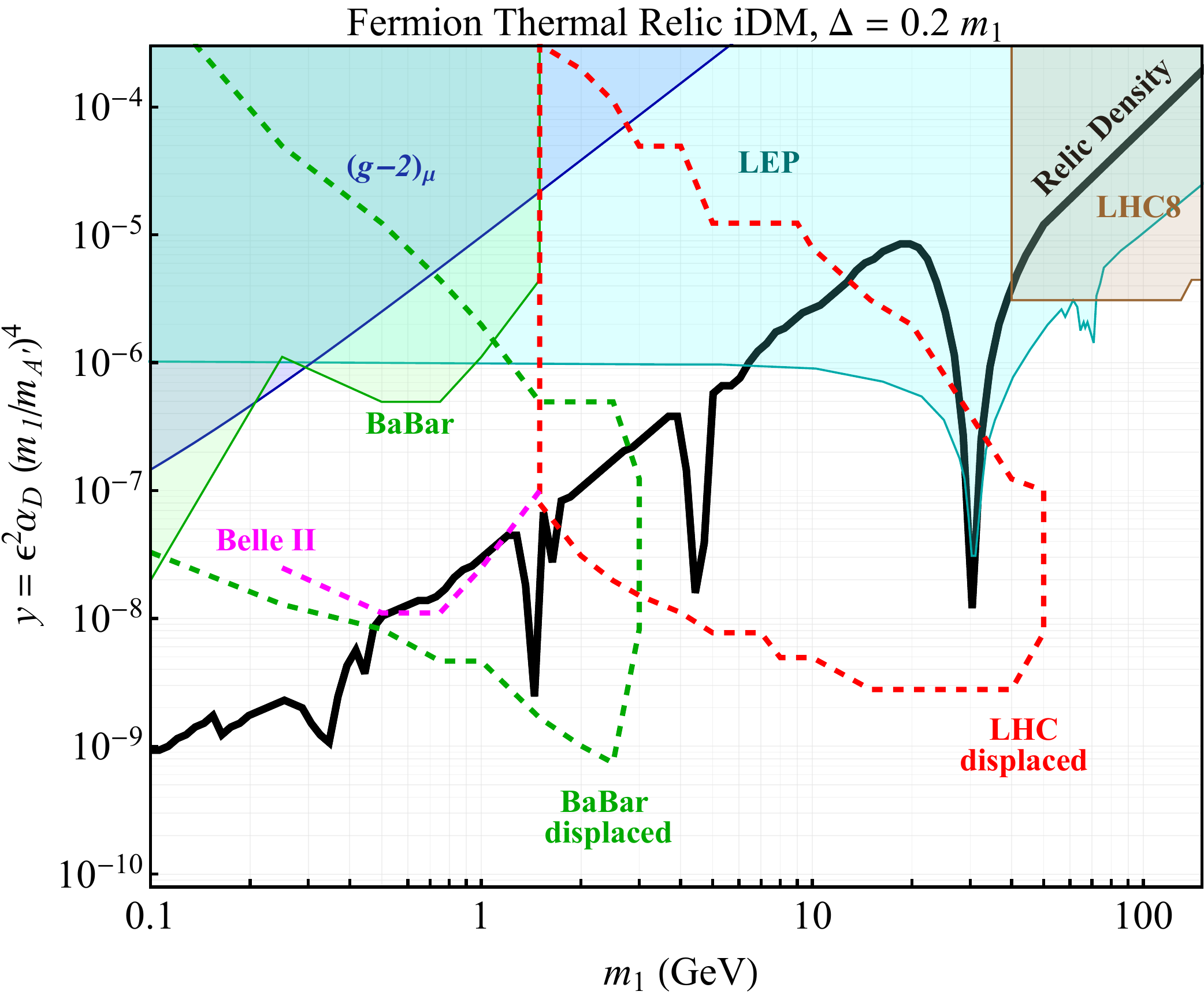}  
  \includegraphics[width=8.6 cm]{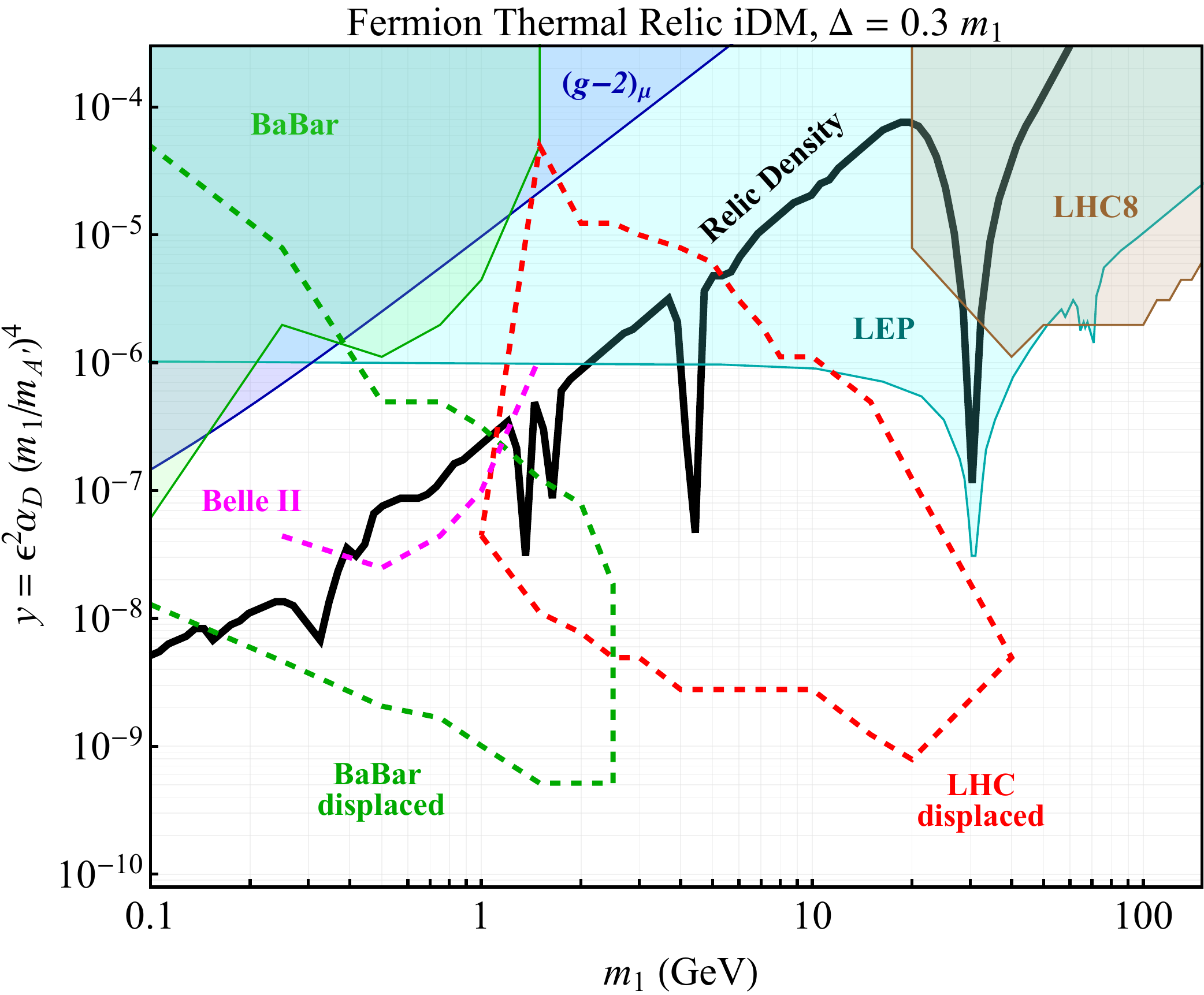}
 \includegraphics[width=8.6 cm]{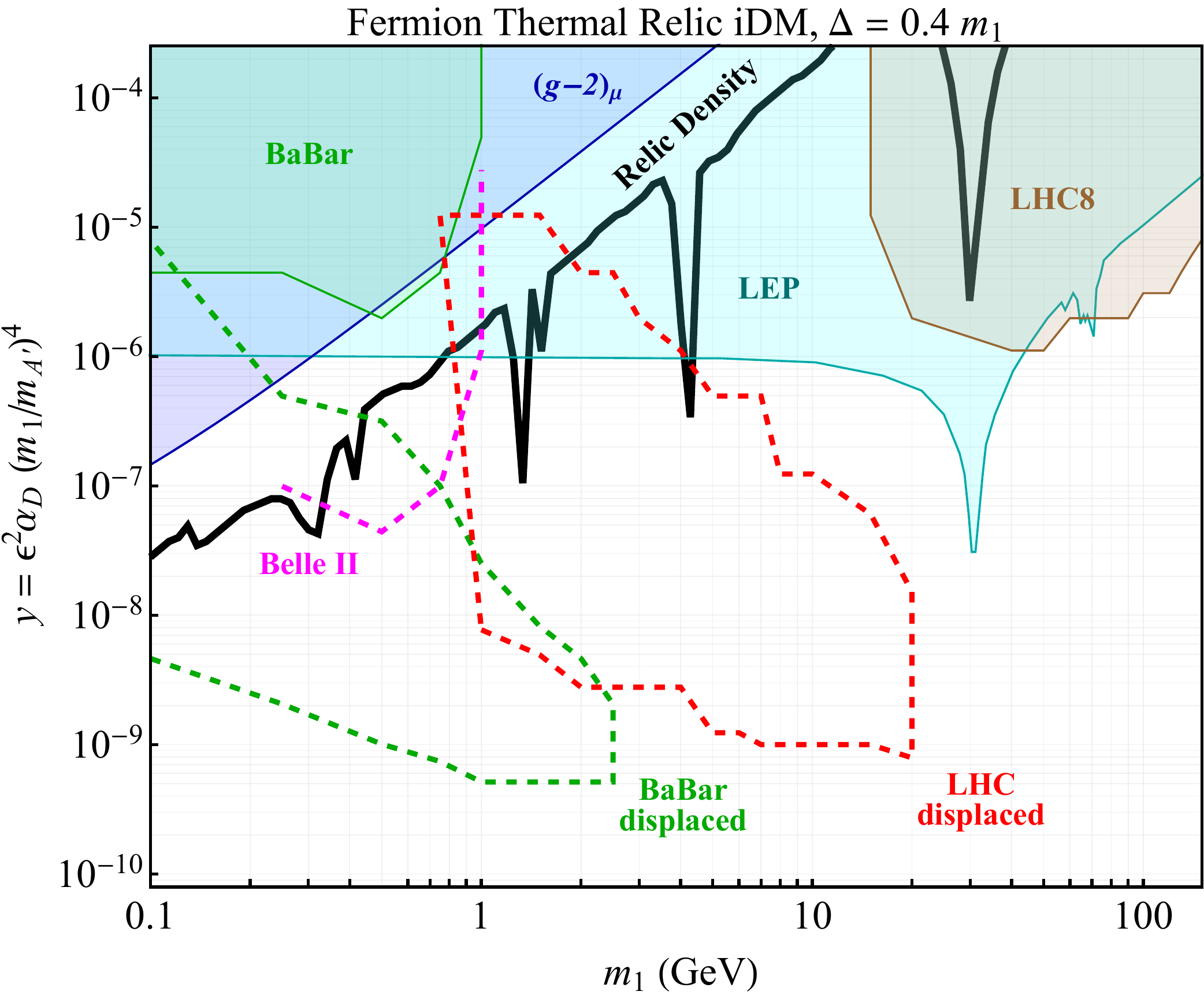}  
\caption{Collider projections for fermionic iDM in the dark photon model with $\alpha_D = 0.1$ and $m_{\apr}/m_1 =3$ vs.~thermal relic density target and other constraints. 
For LHC projections (red dashed), we consider a jet + $\slashed{E}_{\rm T}$ + displaced lepton-jet topology in 13 TeV running with $300\,\,\mathrm{fb}^{-1}$. For $B$-factory projections, we consider existing constraints from BaBar on photon + $\slashed{E}$ (green solid),  projected reach of photon + $\slashed{E}$ + displaced lepton signatures  (green dashed), and projections for a possible Belle II monophoton + $\slashed{E}$ search (purple dashed). See Sec.~\ref{sec:collider-searches} for details. For $\Delta = 0.1m_1$, we also show the projection for a proposed fixed-target missing-momentum experiment (orange dashed) drawn from Ref.~\cite{Izaguirre:2014bca}; since this search would veto visible energy from $\chi_2$ de-excitation, we conservatively  assume it only has sensitivity to $\Delta =0.1m_1$. 
Also shown are constraints from LEP \cite{Hook:2010tw} and $(g-2)_\mu$ \cite{Pospelov:2008zw},
whose sensitivities do not scale with $y$; see Sec.~\ref{sec:constraints}. Both experimental constraints are only sensitive to the visible coupling $\epsilon$ and $m_{\apr}$. To 
avoid overstating these bounds, we conservatively show their $y$ contours for the reasonably large values of $\alpha_D$ and $m_{\apr}/m_1$ given above, which
reveals most of the allowed parameter space (see Sec.~\ref{sec:models}). For smaller values of $\alpha_D (m_1/m_{\apr})^4$, as shown in 
Fig.~\ref{fig:mainplot-fermion-alphaEM}, the $y$-reach for these bounds is greater 
and shifts linearly downwards to cover more of the thermal relic line. The jagged 
spikes represent annihilation to hadronic final states as discussed in Appendix \ref{sec:appendixB-annihilation-rate}.}
\label{fig:mainplot-fermion}
\end{figure*}

\begin{figure*}[t!] 
 \vspace{0.cm}
 \includegraphics[width=8.7cm]{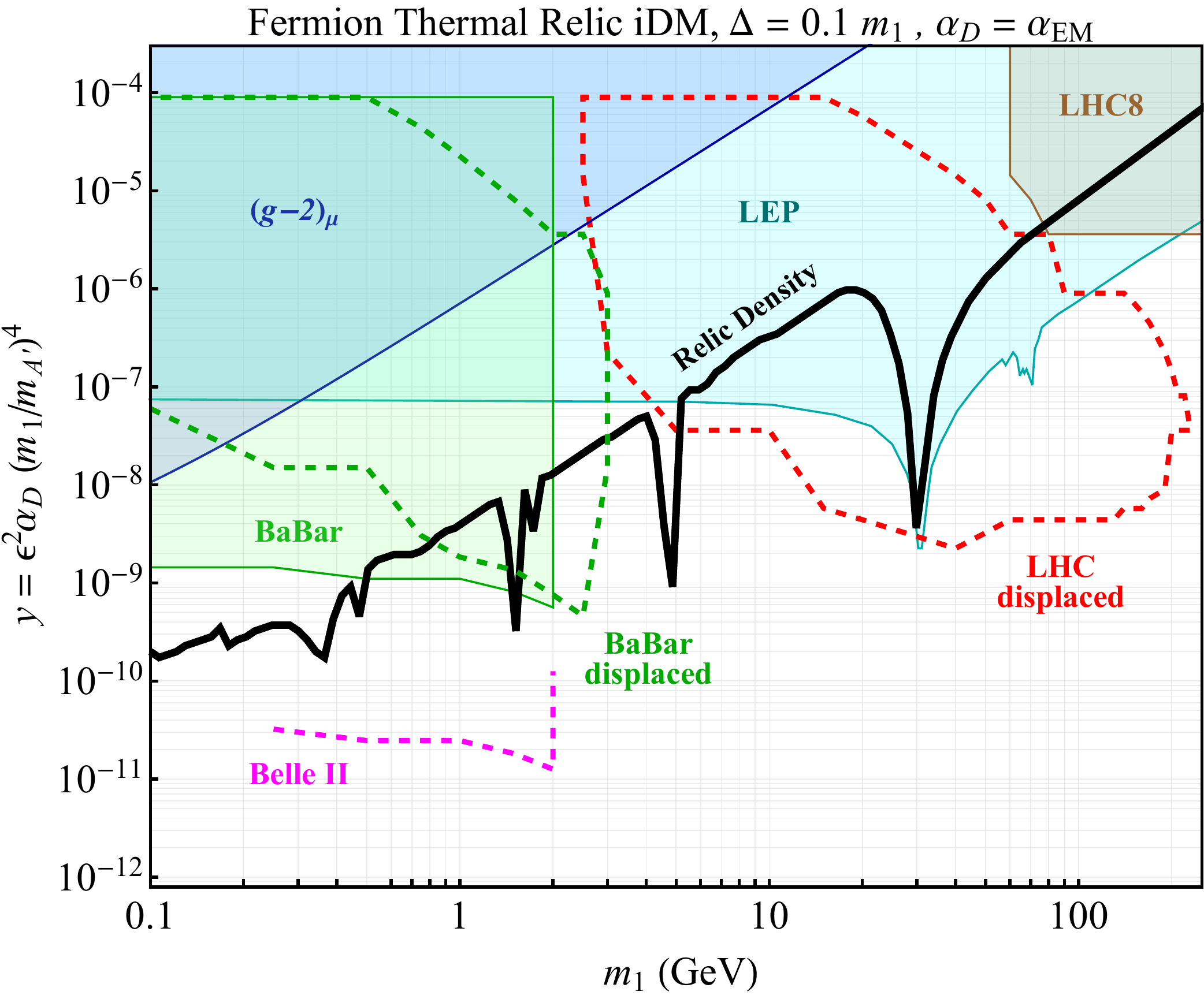}  
 \includegraphics[width=8.7 cm]{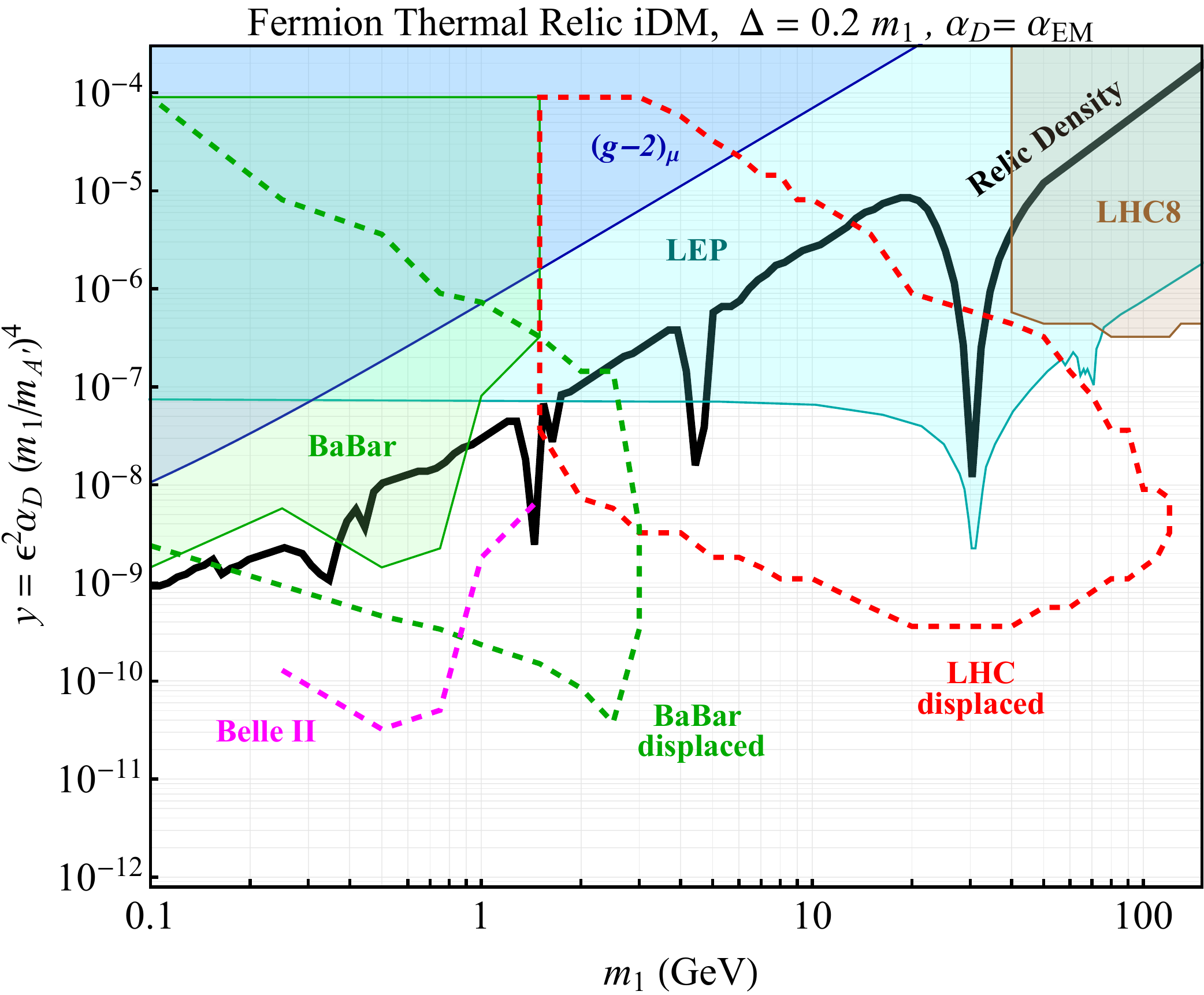}  
 \includegraphics[width=8.7cm]{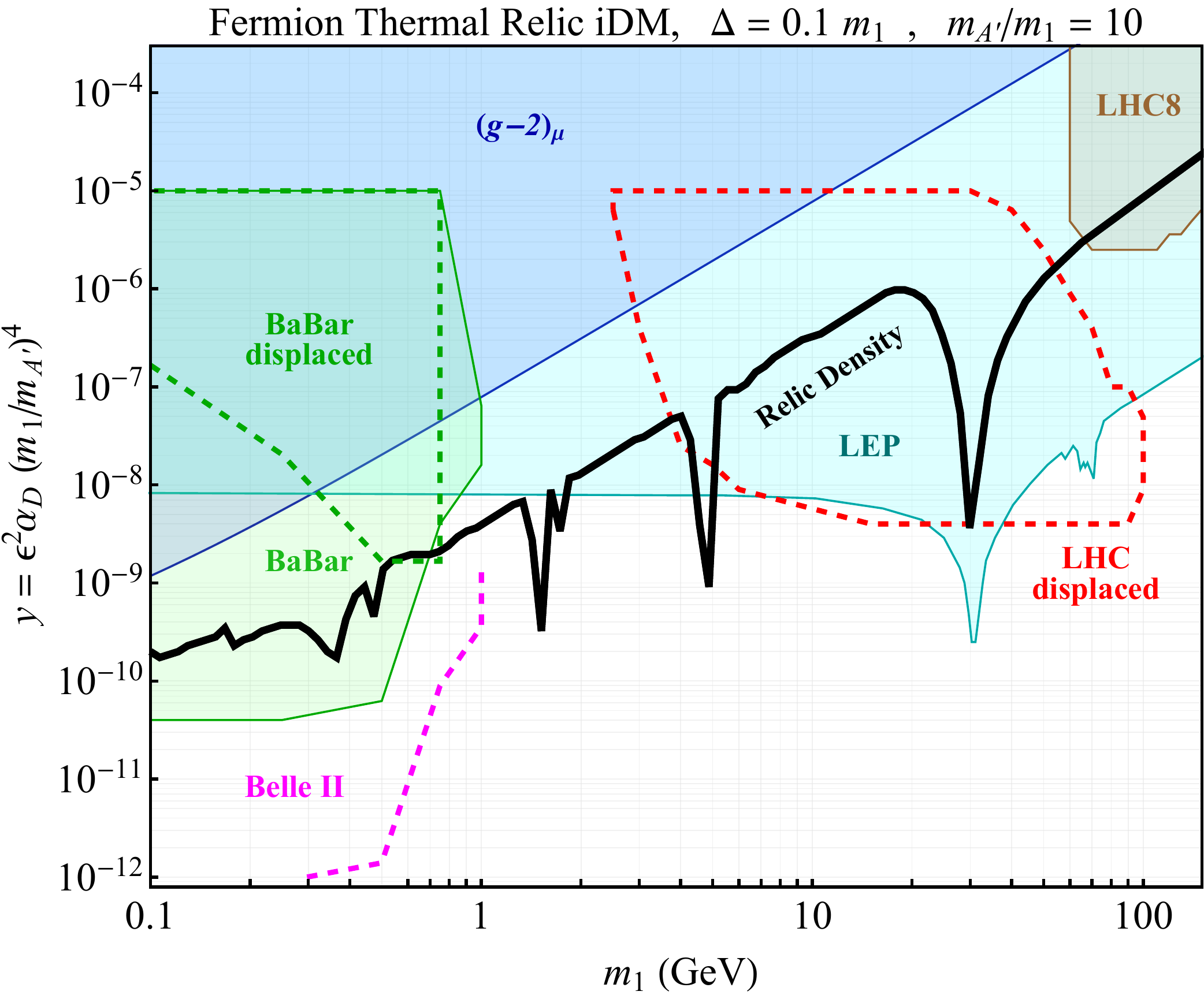}  
 \includegraphics[width=8.7 cm]{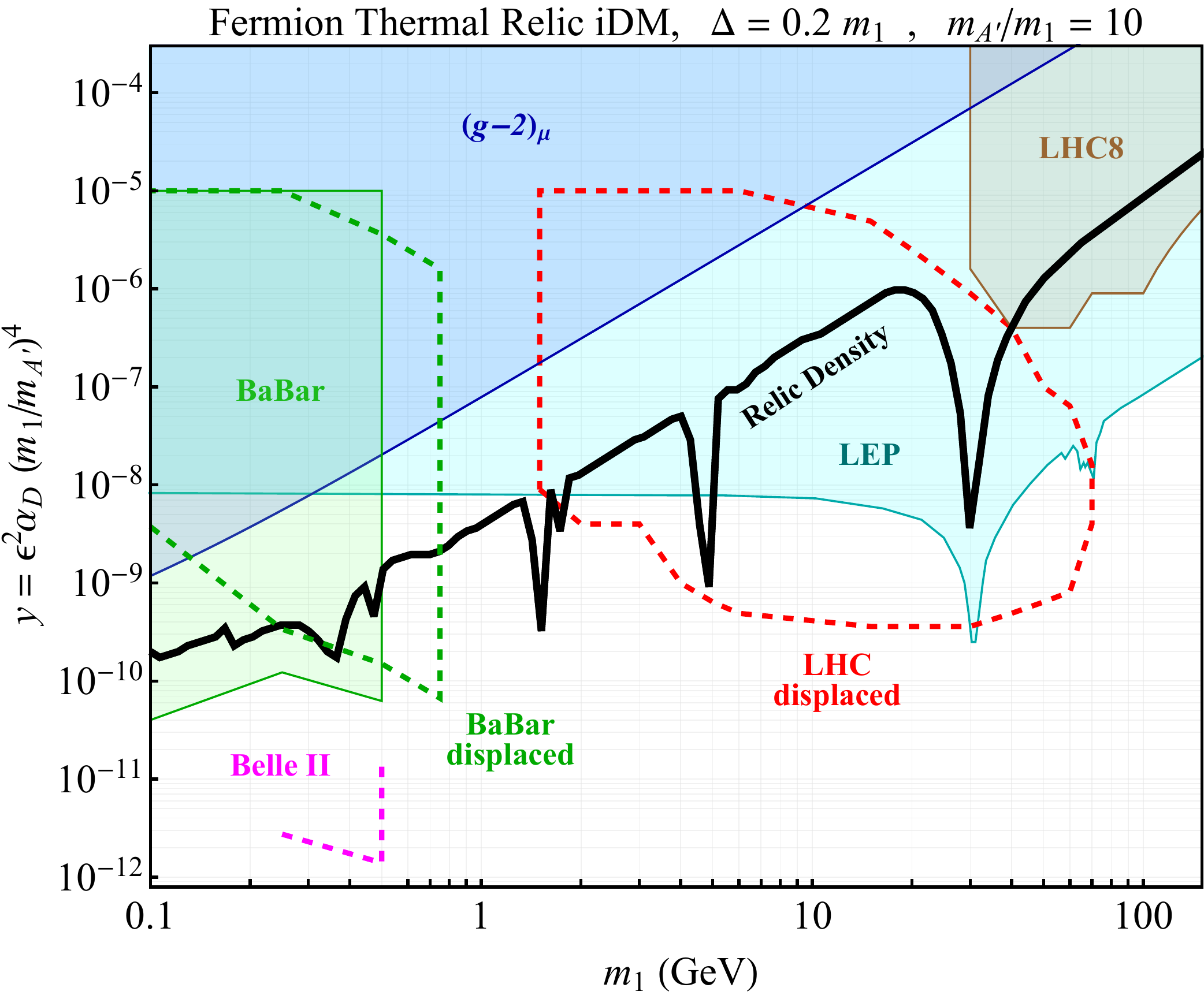}  
\caption{Collider projections for fermionic iDM in the dark photon model with varying $\alpha_D$ and $m_{\apr}/m_1$.
The plots in this figure depict the same mass range and mass splittings as  the top row of Fig.~\ref{fig:mainplot-fermion}.  {\bf Top row:}~same $m_{\apr } / m_1$ ratio as  Fig.~\ref{fig:mainplot-fermion}, but with $\alpha_D = \alpha$ instead of $\alpha_D = 0.1$.  {\bf Bottom row:}~same  $\alpha_D$ as in Fig.~\ref{fig:mainplot-fermion}, but with the DM-mediator mass ratio $m_{\apr } / m_1 = 10$. 
Similar scaling applies to the scalar scenario shown in Fig.~\ref{fig:mainplot-scalar}. 
}
\label{fig:mainplot-fermion-alphaEM}
\end{figure*}

\begin{figure*}[t!] 
 \vspace{0.cm}
 \includegraphics[width=8.6cm]{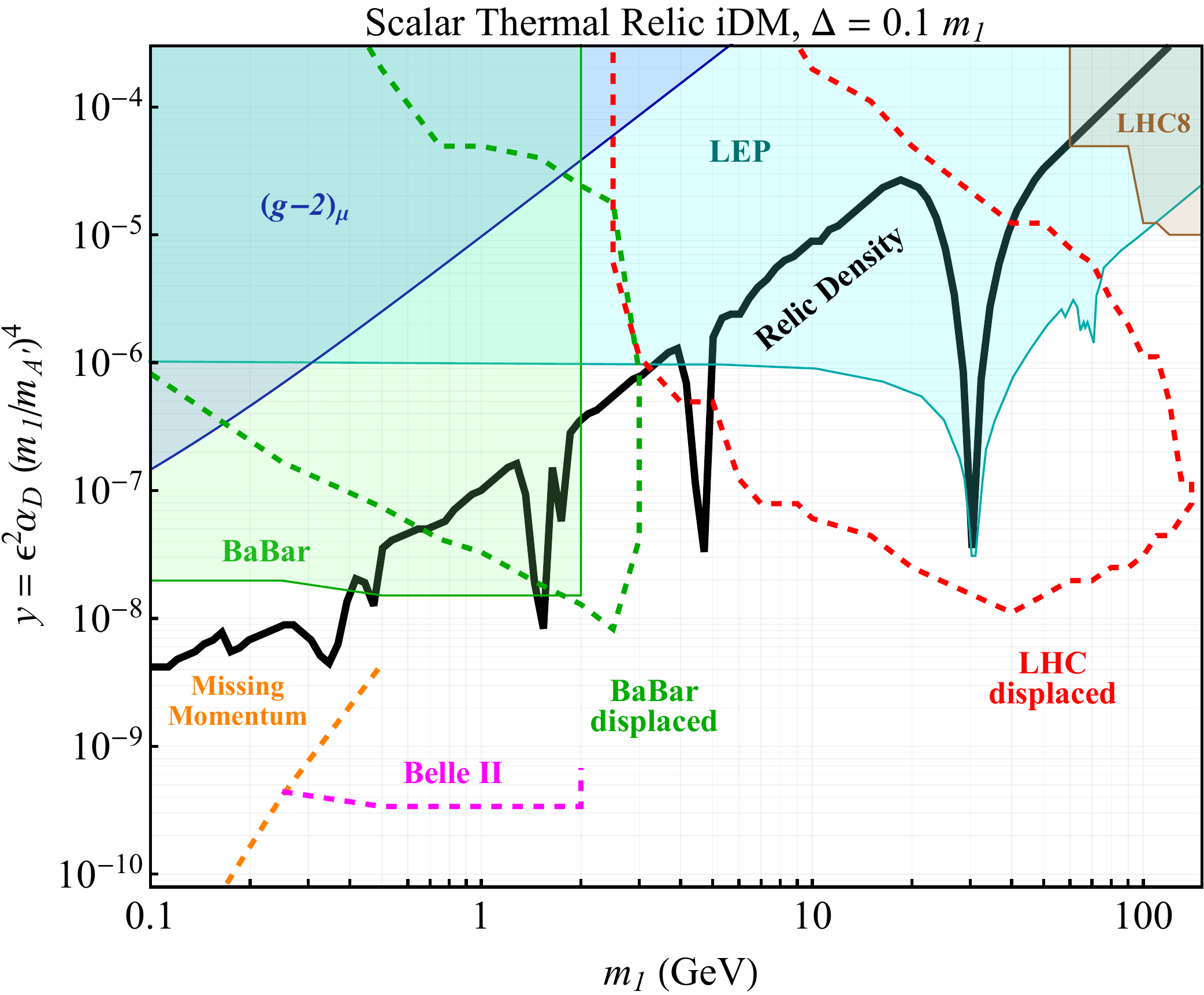}  
 \includegraphics[width=8.6 cm]{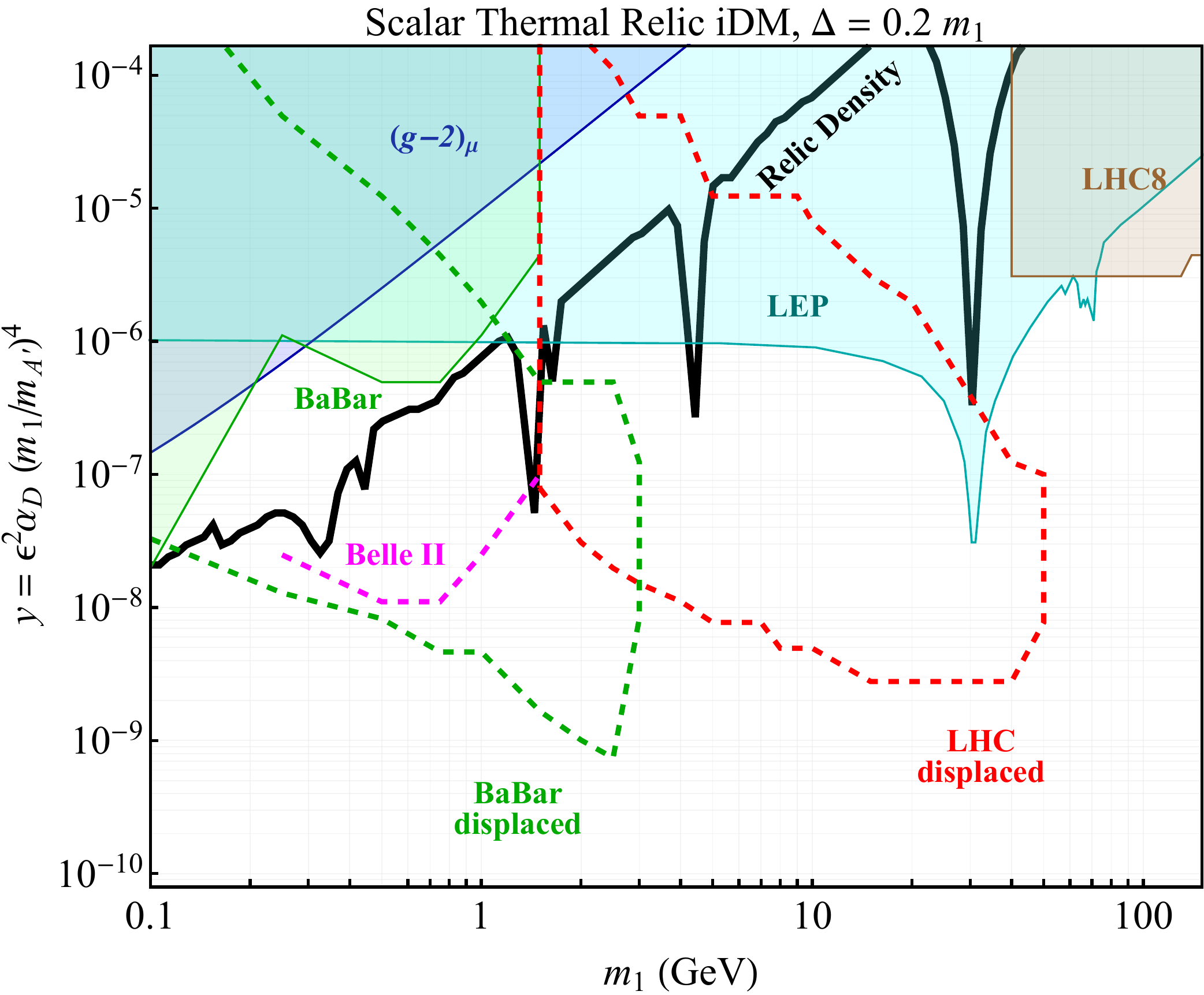}  
  \includegraphics[width=8.6 cm]{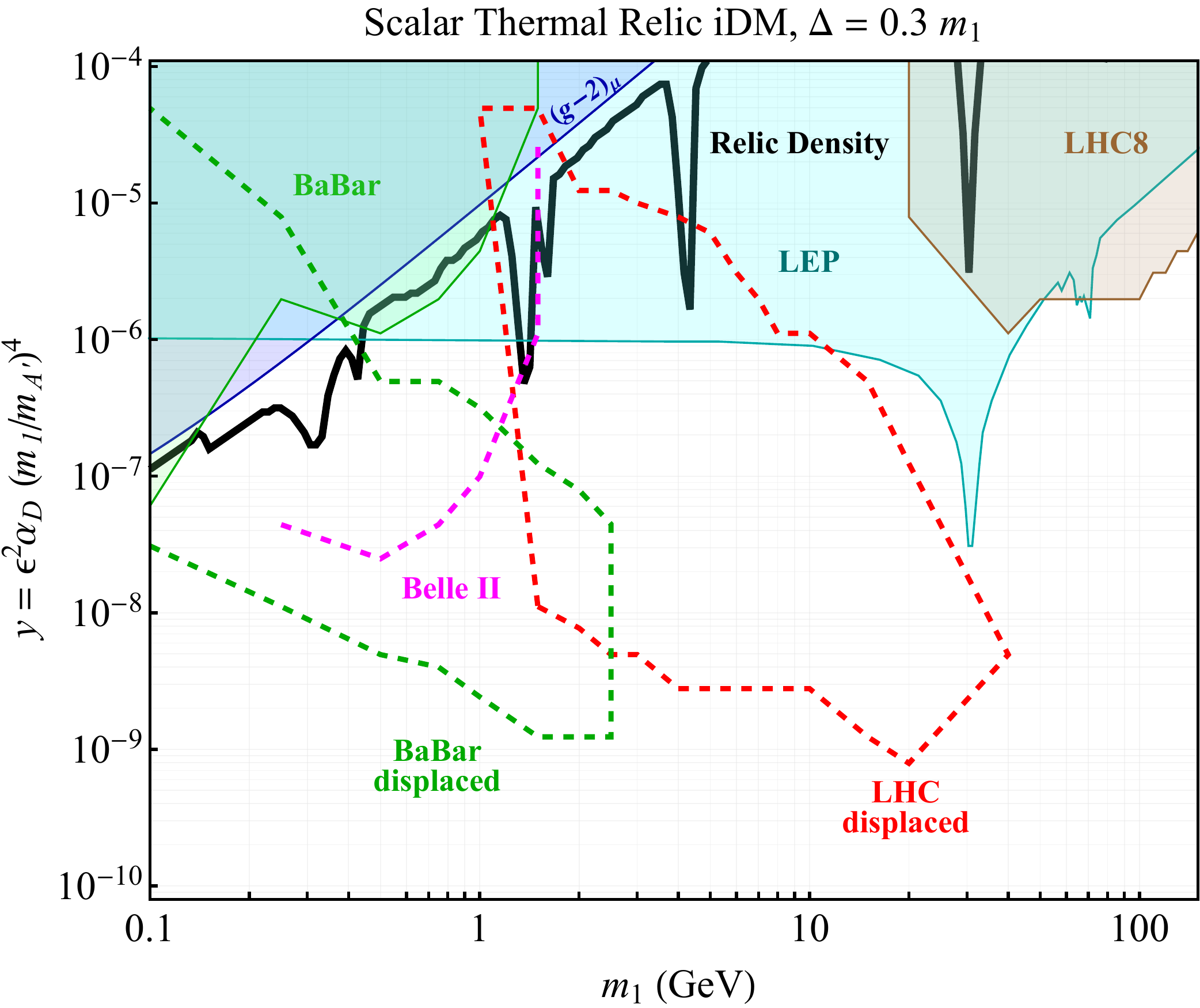}
 \includegraphics[width=8.6 cm]{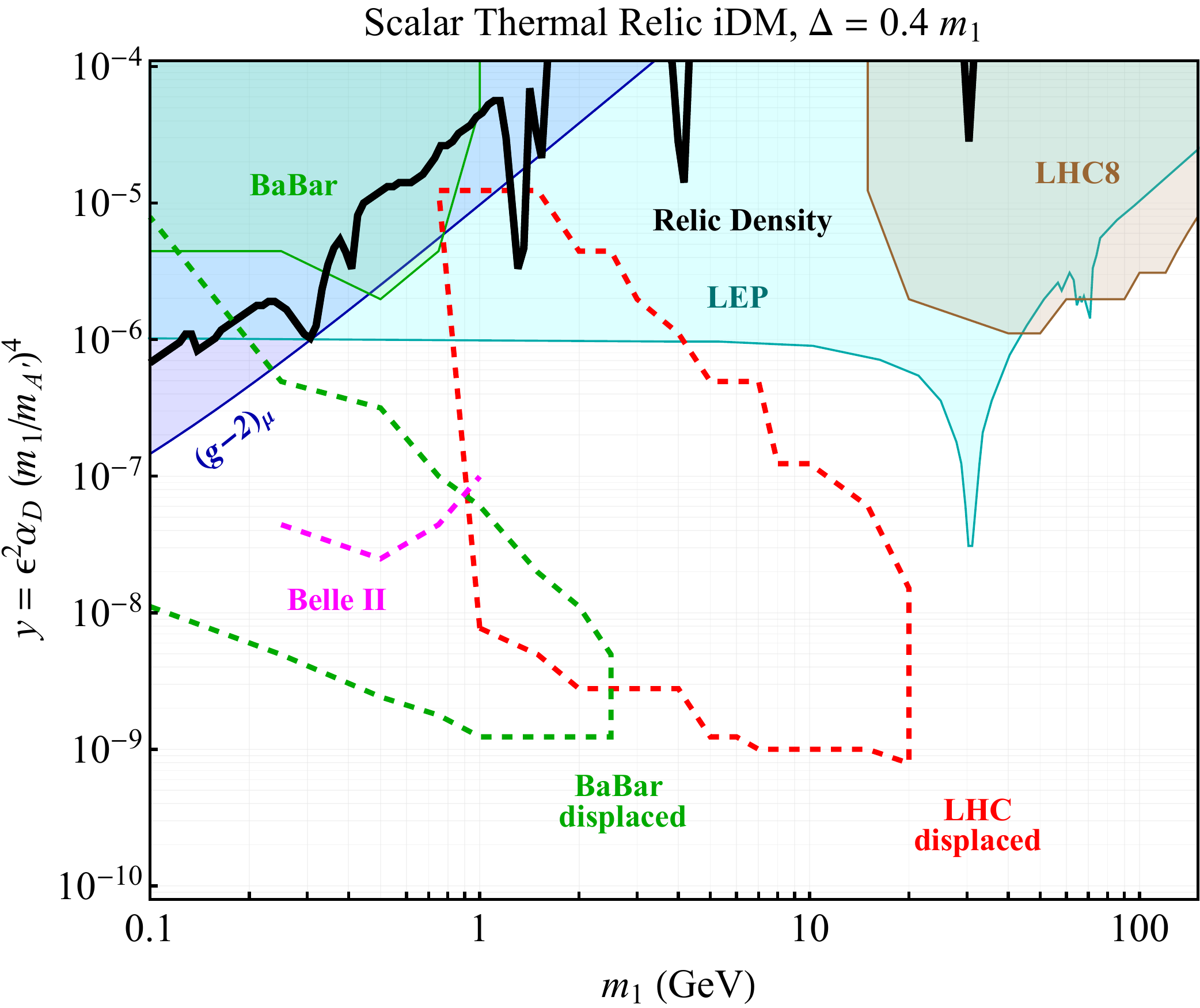}  
\caption{Same as Fig.~\ref{fig:mainplot-fermion} but for scalar iDM in the dark photon model.}
\label{fig:mainplot-scalar}
\end{figure*}

\begin{figure*}[t] 
 \vspace{0.cm}
~~\includegraphics[width=8.6cm]{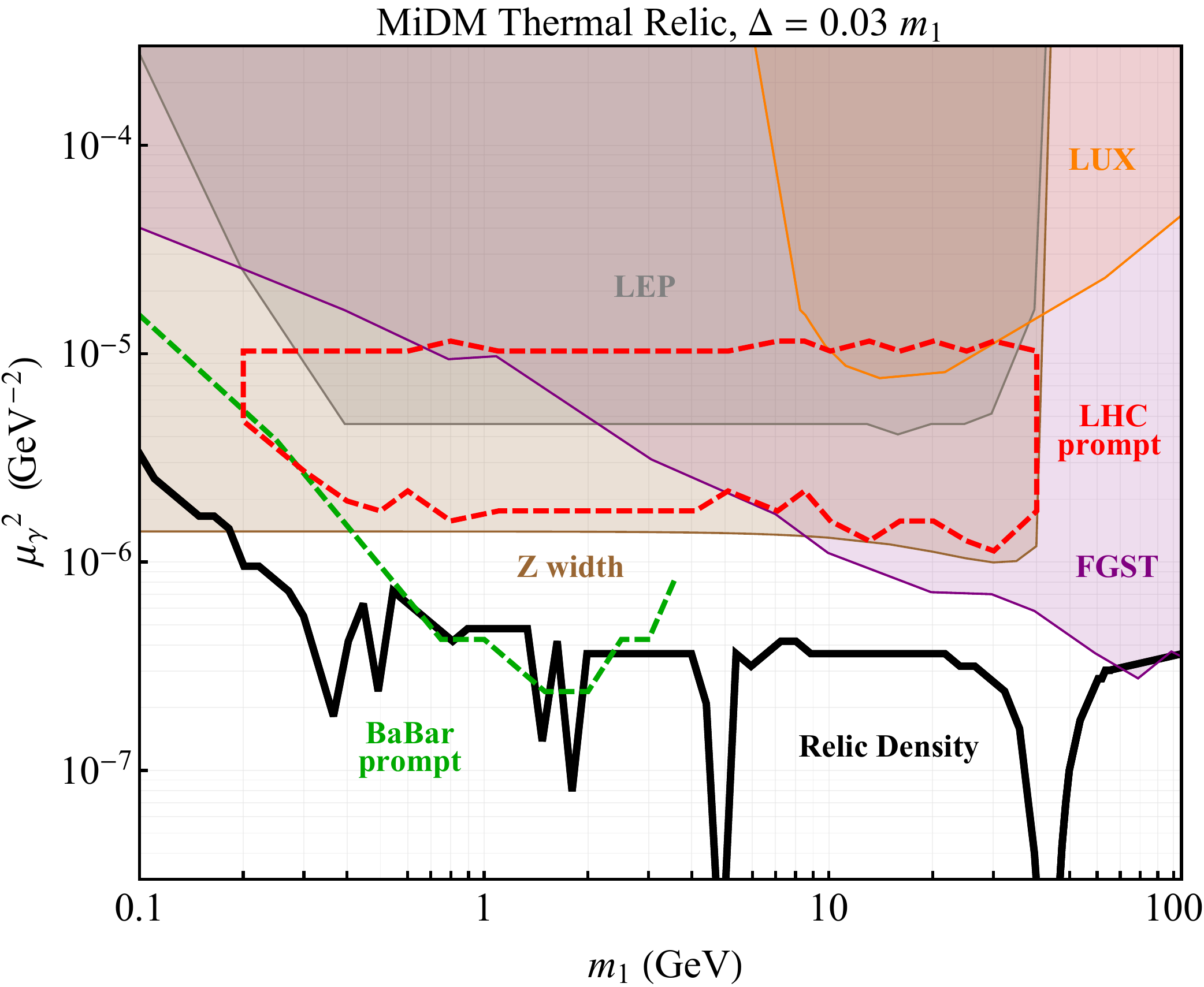}  
~~\includegraphics[width=8.6cm]{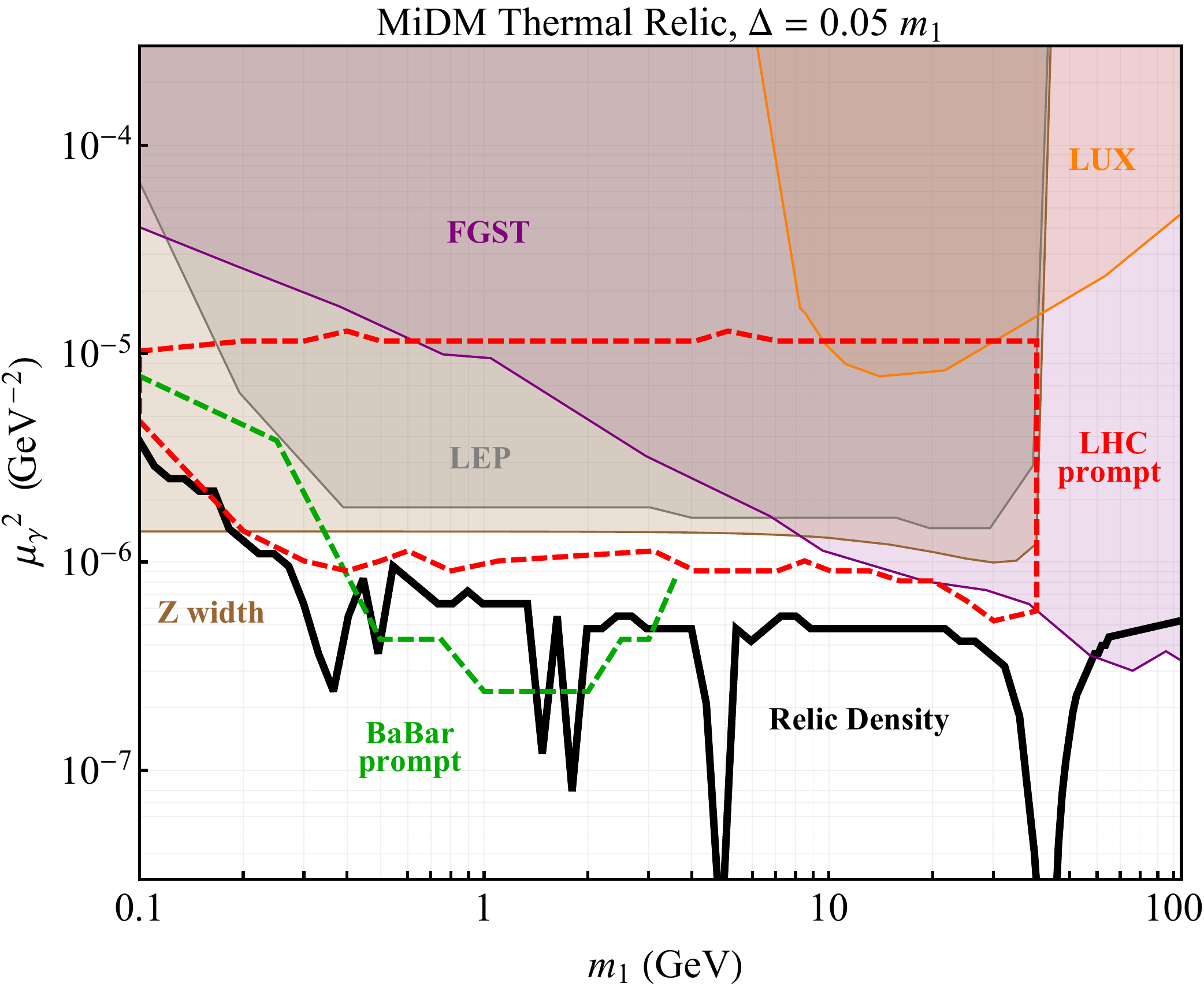}   
~~\includegraphics[width=8.6cm]{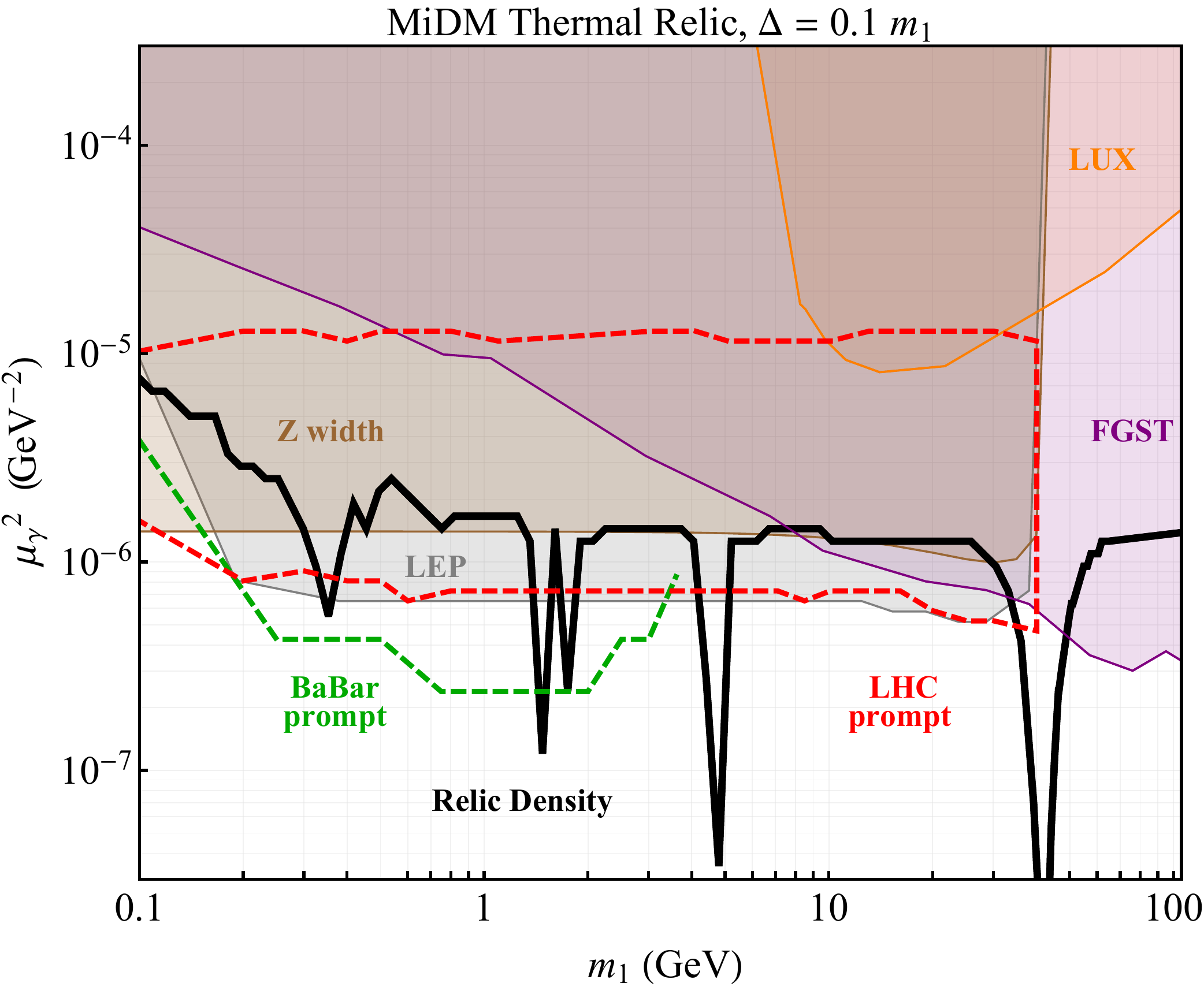}  
~~\includegraphics[width=8.6cm]{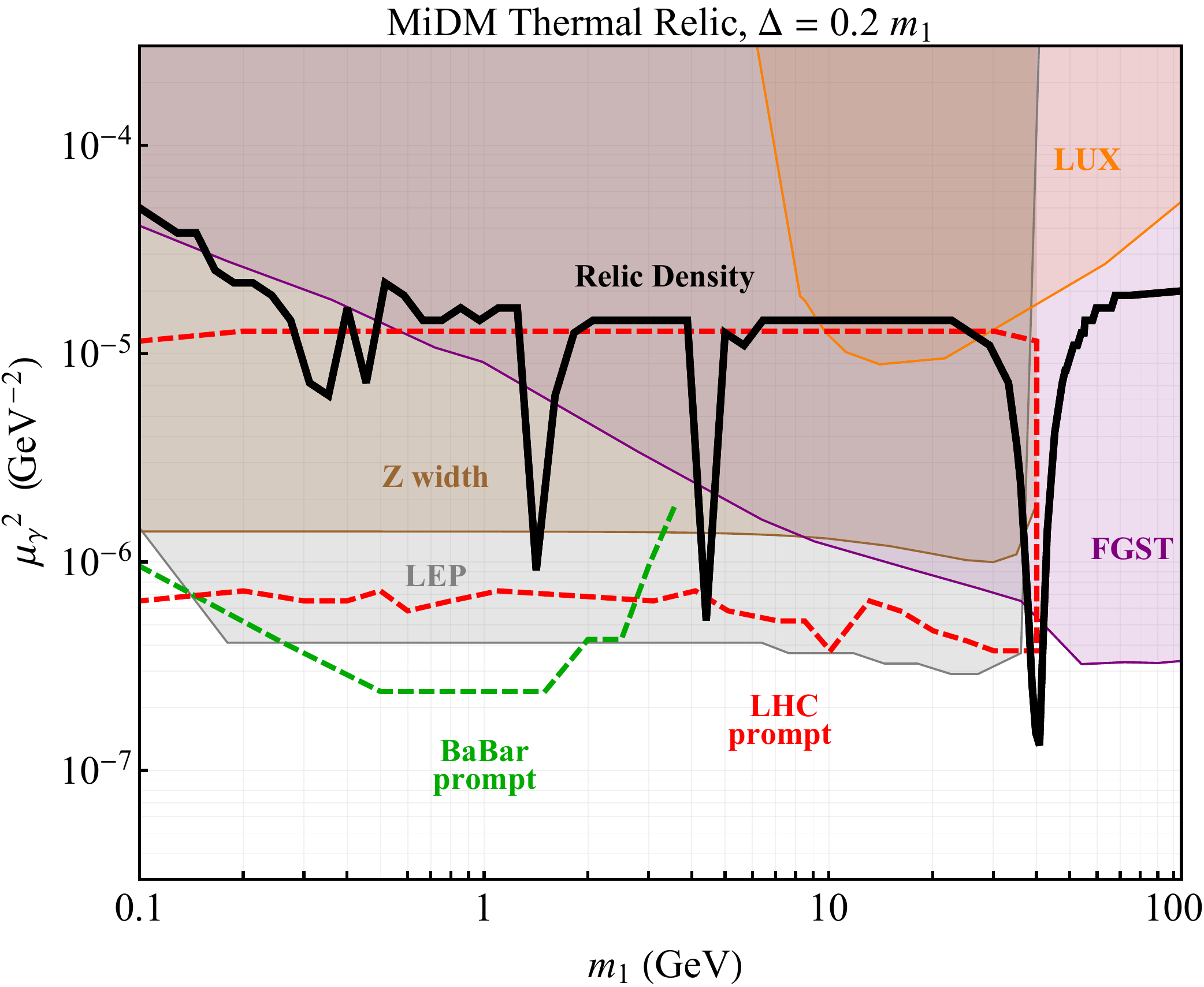}  
\caption{Collider projections for dipole iDM  vs.~thermal relic density target and other constraints. For LHC projections, we consider monojet + $\slashed{E}_{\rm T}$ + photon topology in 13 TeV running with $300\,\,\mathrm{fb}^{-1}$. The curves show projections assuming a prompt  photon. For BaBar projections, we propose a diphoton + $\slashed{E}$ search (green dashed).  See Sec.~\ref{sec:collider-searches} for details.
Also shown are constraints from a LEP diphoton + $\slashed{E}$ search \cite{Acton:1993kp}, LUX \cite{Akerib:2013tjd}, and {\it Fermi} line-searches \cite{Ackermann:2015lka}; see Sec.~\ref{sec:constraints}. We truncate LHC search results where the effective field theory dipole description is no longer valid, which accounts for sharp features at large dipole strength and large mass; see Sec.~\ref{sec:collider-searches} for a detailed discussion.}
\label{fig:mainplot-dipole}
\end{figure*}

In this study, we propose a suite of collider searches for inelastic DM signatures. In particular, we focus on DM and $\mathrm{DM}^*$ masses in the 100 MeV-tens of GeV range, and splittings of order $\sim1-10\%$ of the DM mass, one of the blind spots of the current search program due to the suppression of indirect and direct detection signatures. For such light masses, when the DM and $\mathrm{DM}^*$ states recoil against a comparatively hard jet or photon, the \emph{soft} SM decay products of the excited state are typically aligned with the missing momentum. We show that this feature allows for the effective suppression of the electroweak backgrounds for conventional monojet and monophoton searches. Moreover, for thermal-relic DM-SM couplings and $\mathcal{O}(10\%)$ mass splittings between the ground and excited states, the decay of $\mathrm{DM}^*$ can also occur on macroscopic distances, leading to displaced vertices and other non-prompt phenomena. This results in the possibility of a low-background search over much of the DM parameter space; indeed, the distinctive kinematics of iDM production at colliders allows for sensitivity to the interactions responsible for the cosmological DM abundance. This is in contrast with  traditional collider probes of many elastic DM models, where the large SM backgrounds strongly limit the sensitivity to thermal relic scenarios where the dominant DM-SM interaction is mediated by a new particle with mass at or below the weak scale. We illustrate the sensitivity to iDM in two concrete representative models:~a model where DM interacts with the SM via a kinetically mixed dark photon \cite{Holdom:1985ag}, and a model where DM couples inelastically to the SM   via a magnetic dipole moment \cite{Masso:2009mu,Chang:2010en}. We summarize our results in Figs.~\ref{fig:mainplot-fermion} -- \ref{fig:mainplot-dipole}.

In addition to the iDM signatures considered here, there are many other manifestations of dark sector states at colliders, which can give taggable objects such as hard final-state radiation of new gauge bosons, energetic SM states from excited DM decay, and dark showers \cite{Baumgart:2009tn,Schwaller:2015gea,Cohen:2015toa,Primulando:2015lfa,Autran:2015mfa,Bai:2015nfa,Buschmann:2015awa} that are complementary to our studies.   Inelastic DM decays with monojet + soft hadronic displaced signatures were considered in the contact interaction limit in Ref.~\cite{Bai:2011jg}. Finally, monojet + soft object searches are also useful for compressed supersymmetric spectra (for recent examples, see Ref.~\cite{Giudice:2010wb,Gori:2013ala,Buckley:2013kua,Schwaller:2013baa,Han:2014kaa,Han:2014xoa,Baer:2014kya,Han:2014aea,Barr:2015eva,Nagata:2015pra}), although our work examines parametrically different masses and splittings and different final states, focusing particularly on long-lived decays and exploiting different kinematic features.

 The rest of this paper is organized as follows: in Sec.~\ref{sec:models}, we present the two classes of representative models that this paper studies. We then propose a series of potentially powerful collider searches at both $B$-factories and the LHC in Sec.~\ref{sec:collider-searches}. The cosmology of these models is described in Sec.~\ref{sec:freeze-out}. Finally, we discuss existing constraints on the simplified models in Sec.~\ref{sec:constraints}.


\section{Representative  Models}
\label{sec:models}

The classes of models we consider in this paper all feature dark matter currents that  couple 
{\it inelastically} to the SM. Some of the simplest realizations of DM-SM interactions involve an additional massive
mediator particle that connects SM and DM currents -- {\it e.g.}~a kinetically mixed dark photon or a $Z^\prime$. We also consider the possibility that DM couples to SM gauge bosons via higher-dimensional operators. In the simplest such example, fermionic DM couples to $\gamma/Z$ via a magnetic dipole moment:~if DM is Majorana, this coupling must be inelastic.

We now discuss in turn each of the two models in our study:~a dark photon and magnetic inelastic DM (MiDM).

\medskip

\subsection{Dark Photon Model}

 A simple, well-motivated
 candidate mediator between the dark and visible sector is a massive dark-sector gauge boson $A'$, 
 whose most general renormalizable Lagrangian contains 
\be
 \label{eq:lagrangian}
 \frac{\epsilon_Y}{2} F^\prime_{\mu\nu} B^{\mu \nu} +  \frac{m^2_{\apr}}{2} {\apr}_\mu  {\apr}^\mu  + g B_\mu  {\cal J}_{Y}^\mu + g_D \apr_\mu {\cal J}^\mu_D.
 \ee
Here,  $\epsilon_Y$ is the kinetic mixing parameter, $A^\prime$  is the massive ``dark photon" of a broken $U(1)_D$ symmetry, $B$ is the hypercharge gauge boson, 
 $F^\prime_{\mu\nu}$ 
 and  $B_{\mu\nu}$ are  the dark-photon and hypercharge field strength tensors, ${\cal J}_Y$ is the SM hypercharge current, ${\cal J}_D$ is the dark  current,
  and $m_{{A^\prime}}$ is the dark photon's mass.

After electroweak symmetry breaking, $\apr$ mixes with both $\gamma$ and $Z$, so in the SM mass eigenbasis 
the Lagrangian contains
\be
 \frac{\epsilon_Y}{2} F^\prime_{\mu\nu} B_{\mu \nu} \to  \frac{\epsilon_Y }{2}    F^\prime_{\mu\nu}  \left(          \cos\theta_W  F_{\mu \nu}
 - \sin\theta_W  Z_{\mu \nu} \right) ~.~~
\ee
After diagonalizing the kinetic terms, the  dark photon's couplings to SM fermions are approximately given by \cite{Hoenig:2014dsa}
\be\label{eq:AprimeCoupling}
g_{\apr f\overline f} \approx -\epsilon_Y \frac{m_Z^2 \cos \theta_W e Q_f   - m_{\apr}^2 g Y_f}{m_Z^2 - m_{\apr}^2}~,~
\ee
where $(Y_f) Q_f$ is the SM fermion's (hyper)charge.
In the limit of a light $\apr$, the mixing is predominantly with the photon 
and $g_{\apr \bar f f} \sim  \epsilon_Y \cos\theta_W  e Q_f$,  so the visible sector acquires a millicharge under $U(1)_D$
and we exchange $\epsilon_Y$ for the related parameter
\be
\epsilon \equiv \epsilon_Y \cos\theta_W.
\ee
After $U(1)_D$ symmetry breaking, the DM charge eigenstates will generically mix, giving rise to a split spectrum and inelastic DM, and we show the spectrum and interactions below.

In addition to proposing collider searches for DM coupled via $\apr$, we also  explore how the collider constraints compare to the parameters giving the observed relic abundance and other constraints. Since the $\apr$-mediated scenario depends on five parameters -- the lightest DM mass, $m_1$; the DM mass splitting, $\Delta$; the $A'$ mass, $m_{\apr}$; the dark gauge coupling, $\alpha_D\equiv g_D^2/4\pi$; and $\epsilon$ -- care must be taken to avoid overstating bounds on the parameter space.  In the $\Delta \ll m_1\lesssim m_{\apr}/2$ limit, the DM annihilation  rate largely  depends on only two parameters:~$m_1$, and
 the dimensionless interaction strength, $y$:
 \be
 \sigma v \propto \epsilon^2  \alpha_D \left(\frac{m_1}{m_{\apr}} \right)^4 \equiv y~,~~
 \ee
 which is insensitive to individual choices for each parameter so long as their product remains fixed \cite{Izaguirre:2015yja}. Small values of $\alpha_D$ or of $m_1/m_{\apr}$ would lead to an overabundance of dark matter unless $\epsilon$ is correspondingly larger; on the other hand, different experimental bounds may not scale straightforwardly with $y$. For example, precision QED constraints depend only on $\epsilon$ and $m_{\apr}$, and are independent of $m_1$ and $\alpha_D$. These constraints, expressed in terms of $y$, would therefore be overstated for small values of $\alpha_D$ and $m_1/m_{\apr}$ relative to the $y$ value required for the observed relic abundance. To be  conservative, it suffices
  to choose {\it large}, order-one values of these quantities in computing experimental bounds on $y$. We show later how the results scale for different values of $\alpha_D$ and $m_1/m_{\apr}$.

For  the secluded DM scenario ($m_{\apr}< m_{1}$) \cite{Pospelov:2008zw},  the annihilation process $\mathrm{DM}+ \mathrm{DM} \to \apr+ \apr$ sets the relic abundance, which is independent of the $\apr$ coupling to SM states. Thus,  there is no robust experimental target for this scenario\footnote{For a discussion of the interpolating regime $m_1 < m_{\apr} < 2m_1$, see Refs.~\cite{Izaguirre:2015yja,D'Agnolo:2015koa}.}. It's possible, however, to still produce $\mathrm{DM}+\mathrm{DM}^*$ through a virtual $A'$ at colliders, with subsequent decay $\mathrm{DM}^* \rightarrow \mathrm{DM} +A'$. The $A'$ subsequently decays into SM final states. This kind of topology would fall under the scenarios studied by Refs.~\cite{Bai:2015nfa,Autran:2015mfa}.  Furthermore, the direct production of the DM states may not be the discovery channel of this class of models, as now the $A'$ could be produced directly and observed through its decays into $\ell^{+}\ell^{-}$ or into dijets (see Refs.~\cite{Hoenig:2014dsa,Curtin:2014cca} for recent studies). Because $m_{\apr}\gtrsim 2m_1$ offers a clear, experimentally promising target for the parameters giving the observed relic abundance, we focus on that scenario.\\

Returning to iDM,  the  $y$ necessary for freeze-out grows with increasing mass splittings, $\Delta$, but is still a useful variable to characterize the parameter space for fixed $\Delta$.  
For a purely inelastic coupling, the $\Delta \gsim m_1$ regime is excluded by a combination of collider and precision-QED 
 probes (see Figs.~\ref{fig:mainplot-fermion}--\ref{fig:mainplot-scalar}). Similarly, for sufficiently small DM parameters
  $m_1/m_{\apr} $ or $\alpha_D \ll 1$, these same constraints rule out the thermal freeze-out hypothesis (see Fig.~\ref{fig:mainplot-fermion-alphaEM}). 
Thus, the viable parameter space for thermal iDM coupled to an $\apr$ 
requires $ \Delta \lsim \mathcal{O}(10 \%)$, comparable DM/mediator 
masses, and sizeable $\alpha_D \lsim 1$, 
so our search strategy in this paper  primarily targets this regime.

\subsubsection*{  Inelastic Fermion Current}

\noindent Consider a familiar Dirac spinor $\psi = (\eta~~ \xi^\dagger)$ charged under the  $\mathrm{U}(1)_D$ gauge symmetry. The vector current is
 \be
{\cal  J}^\mu = \overline \psi \gamma^\mu \psi =   \eta^\dagger \overline \sigma^\mu \eta - \xi^\dagger \overline\sigma^\mu \xi ,
 \ee
where $\eta$ and $\xi$ are two-component Weyl fermions. Gauge invariance allows only a Dirac mass term, $m_D$. However, when the symmetry is spontaneously broken, the components of $\psi$ can also acquire Majorana masses:
\be   
 \hspace{-0.2cm }-{\cal L} \supset m_D\eta \xi +\frac{m_\eta}{2} \eta \eta \! + \frac{m_\xi}{2} \xi\xi   + \mathrm{h.c.},~
\ee
Because the $\mathrm{U}(1)_D$ symmetry is restored in the limit that the Majorana masses go to zero, it is natural for the Majorana masses to be smaller than the Dirac mass. In this limit, the spectrum is split in the diagonal mass basis into two nearly equal Majorana mass eigenstates, where the mass eigenstates couple predominantly inelastically. For example, if  the Majorana masses are equal, 
($m_\eta = m_\xi$), the
mass eigenstates are $\chi_1 = i(\eta - \xi)/\sqrt{2} ~,~\chi_2 = (\eta + \xi)/\sqrt{2}$
with eigenvalues $m_{1,2} = m_D \mp m_M$, and the vector current now couples different states to one other,
\be\label{eq:fermion-current}
{\cal J}^\mu  =       i (\chi_1^\dagger \overline \sigma^\mu \chi_2 - \chi_2^\dagger \overline \sigma^\mu \chi_1 ) \equiv {\cal J}^\mu_{\rm iDM} ~,
\ee
where the iDM subscript emphasizes the inelasticity of the interaction. 

In the more general case where $m_\eta \ne m_\xi$, the mass eigenvalues are 
\be
m_{1,2} = \sqrt{m_D^2+(m_\eta-m_\xi)^2/4}\pm(m_\eta+m_\xi)/2~,~~
\ee
 and the physical splitting is $\Delta=m_\eta+m_\xi$. In this case, the vector current contains an additional elastic piece:
\be \label{eq:general-splitting}
{\cal J}^\mu &=&\frac{m_D}{\sqrt{m_D^2+(m_\xi-m_\eta)^2/4}}\,{\cal J}_{\rm iDM}^\mu   \nonumber\\
&&{} +\frac{m_\xi-m_\eta}{\sqrt{4m_D^2+(m_\xi-m_\eta)^2}} ( \chi_2^\dagger \overline \sigma^\mu \chi_2 -
\chi_1^\dagger \overline \sigma^\mu \chi_1    )~.~~~~~~
\ee
Because we  consider cases where $\Delta$ is not too much smaller than the the Dirac mass, we include the effects of the elastic coupling by assuming $m_\eta = \Delta$, $m_\xi=0$ when determining various constraints on the model.

\medskip
\subsubsection*{  Inelastic Scalar Current} 
\medskip

\noindent Similar conclusions follow if the DM is a complex scalar $\varphi$ with vector current
\be
{\cal J}^\mu = i (  \varphi^* \partial^\mu  \varphi -  \varphi \partial^\mu  \varphi^* )~.~~
\ee
If the $\varphi$ scalar potential contains both $U(1)_D$-preserving and -violating mass terms,  the  l mass terms for $\varphi$ can be written as
\be \label{eq;scalar-mass}
-{\cal L} \supset  \mu^2 \varphi^* \varphi  + \frac{1}{2} \rho^2 \varphi \varphi +\mathrm{h.c.},
\ee
where $\mu$ and $\rho$ are ``Dirac" and ``Majorana"-like mass terms for a scalar particle, which
respectively preserve and break any gauge symmetry under which $\varphi$ is charged\footnote{A mass term of type $V(\varphi) = (\rho')^2\mathrm{Im}(\varphi^2)/2$ is also allowed; however, because the scalars are real in the mass basis, an $\mathrm{SO}(2)$ rotation among the scalars does not induce a diagonal coupling between $\apr$ and the mass eigenstates, and so it has no effect on our results.}. We have assumed
that the $\varphi$ expectation value remains zero in order to preserve a symmetry to stabilize the DM. 

Diagonalizing  the mass terms in 
Eq.~(\ref{eq;scalar-mass}) yields eigenstates  $\phi_{1,2}  = (\varphi \pm  \varphi^*)/\sqrt{2}$ with
corresponding mass-squared eigenvalues $\mu^2 \pm \rho^2$. The vector current is now
\be\label{eq:scalar-current}
{\cal J^\mu} =  i(   \phi_1 \partial^\mu \phi_2   -  \phi_1 \partial^\mu \phi_2  )~,~
\ee
which is purely off-diagonal. Because the scalar mass eigenstates are real, no diagonal interactions with a single $\apr$ are allowed. The 
covariant derivative also yields diagonal quartic terms of the form $\phi_i \phi_i \apr \apr$, which
 gives an elastic scattering mode involving pairs of gauge bosons,  but the coupling to fermions in colliders and other experiments is suppressed by one loop.


\begin{figure}[t!]
 \vspace{0.5cm}
  \hspace{0.5cm}
\includegraphics[width=6cm]{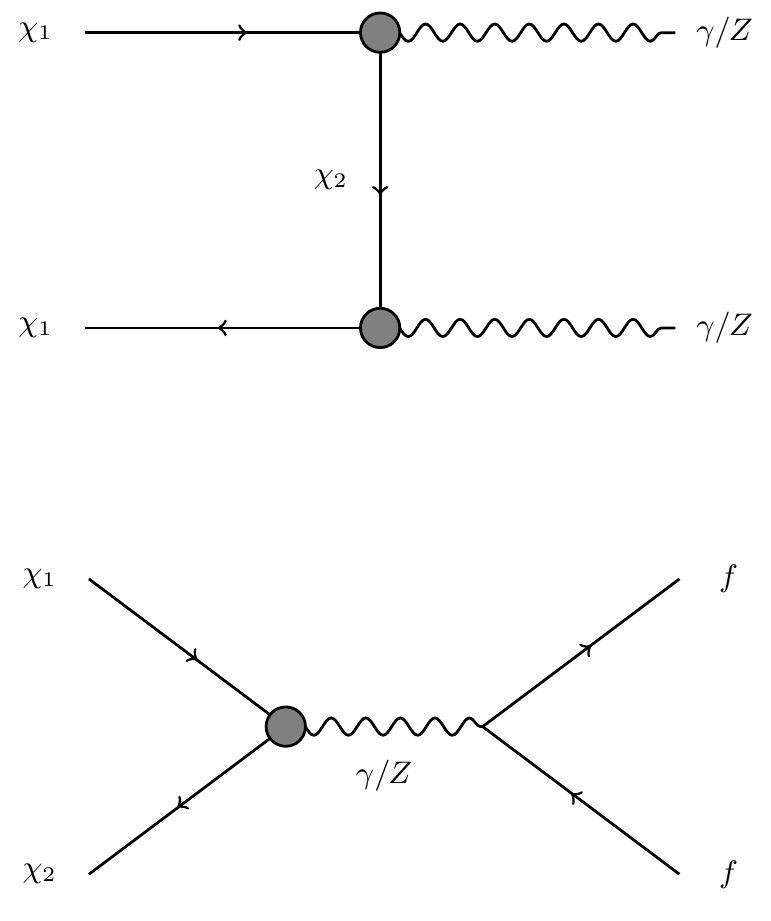}  ~
\caption{Leading self-annihilation (top) and inelastic annihilation (bottom)  diagrams through the effective dipole interaction depicted as a 
solid dot. }
\label{fig:DM_dip_ann}
\end{figure}

 \subsection{$\gamma/Z$-Mediated Dipole Interaction}
The other representative model we consider is magnetic inelastic DM (MiDM) \cite{Masso:2009mu,Chang:2010en}. If we consider a DM Dirac  fermion $\psi = (\eta~ \xi^\dagger)$,
it can have a direct coupling to the SM through the dipole interaction
\be
&& ~~~~~~ {\cal L} =
 \frac{ \mu_B}{2} \,  \overline \psi \, \sigma_{\mu \nu} B^{\mu \nu} \psi ~,~~  \\ \nonumber
\\
&=&\frac{ i \mu_B}{4}       \left( \xi \sigma^{[\mu\,,} \overline \sigma^{\nu]} \eta + \xi^\dagger \overline \sigma^{[\mu\, ,} \sigma^{\nu]} \eta^\dagger \right)     B^{\mu \nu} + \mathrm{h.c.}~ ,~~~
\ee 
where $\mu_B$ is a dipole moment of dimension inverse mass,  $\sigma^{\mu \nu} = \frac{i}{2}[\gamma^\mu , \gamma^\nu]$ and $\sigma_\mu  (\overline \sigma_\mu) = ({ \mathbf1},  \pm \sigma_i)$, where
$\sigma_i$ is a Pauli matrix. This interaction is well-motivated and arises in a wide variety of   
DM models \cite{Pospelov:2000bq,Pierce:2014spa,Fortin:2011hv,Barger:2012pf,Barger:2010gv,Banks:2010eh,Heo:2009vt,Sigurdson:2004zp,Weiner:2012cb}.

In the absence of an exact DM $\mathrm{U}(1)$ symmetry, the components of $\psi$ can also have Majorana masses. In terms of the mass eigenstates,  $\xi (\eta) = (\chi_1 \pm i\chi_2)/\sqrt{2}$,  this operator becomes
 \be
 \frac{ \mu_B}{2}  \overline \chi_1 \sigma_{\mu \nu} B^{\mu \nu} \chi_2  + \mathrm{h.c.}~,~~ 
 \ee
 where we have built 4-component Majorana fermions from each of the $\chi_{1,2}$ Weyl spinors. 
For the dipole-mediated scenario, we  assume this operator is the lowest-dimension interaction arising from 
a UV theory with additional (hyper) charged field content; for an explicit construction,
 see Ref.~\cite{Weiner:2012cb}. 
 After electroweak symmetry breaking, this operator decomposes into  
\be
{\cal L} = \frac{ \mu_{\gamma}}{2} \,
 \overline \chi_1 \sigma_{\mu \nu} F^{\mu \nu} \chi_2 + \frac{ \mu_{Z}}{2}   \overline \chi_1 \sigma_{\mu \nu} Z^{\mu \nu} \chi_2  + \mathrm{h.c.} ~,~~
\ee 
in the  $\gamma$ and $Z$ mass eigenbasis, where $\mu_Z / \mu_\gamma = - \tan \theta_W$; although
 higher-dimensional operators can modify this ratio \cite{Weiner:2012cb}, we consider the simplest case where the ratio is given by the weak mixing angle, and we parameterize our results in terms of $\mu_\gamma$. Because dipole moments vanish for Majorana fermions, there is no diagonal coupling akin to the interactions of fermion iDM with a dark photon. However, in computing DM annihilation rates, we include the effects of the higher-order $t$-channel processes $\chi_1\chi_1\rightarrow \gamma\gamma$, $\gamma Z$, and $ZZ$ (see Fig.~\ref{fig:DM_dip_ann}).
 
 As in the $\apr$ scenario, our focus is on covering the parameter
 space that induces a thermal-relic annihilation in the early universe. However, a crucial difference in this case is that, for a given  $\Delta$,  
 the annihilation rate depends only on $\mu_\gamma$ and  $m_1$, so no additional assumptions need to be made in order
 to compare different kinds of experimental bounds against the thermal relic benchmark. Thermal Majorana MiDM with a mass splitting $\Delta \gsim 0.2\, m_1$ 
 is nearly excluded already by a combination of direct, indirect, and collider searches (see Fig.~\ref{fig:mainplot-dipole}). For 
smaller mass splittings, only the $ m_1 \gsim 100~ \GeV$ region is robustly ruled out by gamma-ray line searches, so the searches proposed in this paper
are designed to target the remaining viable parameter space.

%

\section{Collider Search Proposals}
\label{sec:collider-searches}

Dark matter searches at high-energy colliders traditionally feature missing (transverse) energy ($\slashed{E}_T$ or MET) and fit into two broad categories: searches for DM produced from the decays of additional new SM-charged states such as $t$-channel mediators \cite{Papucci:2014iwa,Garny:2015wea,An:2013xka},
and searches targeting direct DM production through the reaction $pp\rightarrow \rm{DM}+\rm{DM} + {\it X}$ where $X$ is some visible SM state. The former class is more model-dependent by nature, although well-motivated frameworks like supersymmetry (SUSY) fall into this category; by contrast, the latter is more model-independent  because it relies primarily  on DM's direct coupling to the SM. In recent years, there have been many proposed  searches and analyses for DM pair-production, which yields $\slashed{E}_T$ in association with SM final states, including monojet, monophoton, and mono-boson  \cite{Petriello:2008pu,Gershtein:2008bf,Cao:2009uw,Beltran:2010ww, Bai:2010hh,Fox:2011fx,Fox:2011pm,Rajaraman:2011wf, Bai:2011jg,Goodman:2010ku,Goodman:2010yf,Bai:2012xg,Carpenter:2012rg,Lin:2013sca,Aad:2013oja,Carpenter:2013xra,Khachatryan:2014rwa,Aad:2014tda,Berlin:2014cfa, Aad:2014vka,ATLAS:2014wra,Khachatryan:2014tva,Izaguirre:2014vva,Khachatryan:2014rra,Aad:2014vea,Aad:2015zva,Haisch:2015ioa}. Indeed, the LHC is particularly well-suited for discovering classes of DM models with  contact interactions of light DM, where the sensitivity of direct and indirect-detection experiments is sub-optimal, although these searches remain insensitive to  contact-interaction strengths sufficient to induce thermal-relic annihilation rates  
in the early universe for light mediators.

In an extended DS, two different particles in the DS can be produced in association at colliders, in contrast with mono-$X$ searches that target only the production of ground-state particles. As a result, we show that the collider sensitivity to scenarios such as iDM can be enhanced by tagging the decay products of the associated state(s), providing a powerful handle for background rejection. We focus on the representative models in Sec.~\ref{sec:models}. However, there also exist  models where the DM lives in an extended DS, which can give rise to more varied and spectacular signatures than those we consider (see, for example, Refs.~\cite{Baumgart:2009tn,Schwaller:2015gea,Cohen:2015toa,Primulando:2015lfa,Autran:2015mfa,Bai:2015nfa,Buschmann:2015awa}).

We now  propose a series of new searches at both the LHC and $B$-factories that can dramatically improve the sensitivity to the scenarios introduced in Sec.~\ref{sec:models}. We organize our searches by model, since the models give very different signatures at colliders.

\subsection{Dark photon}

\subsubsection*{LHC}

\begin{figure}[t] 
 ~~\includegraphics[width=8.5cm]{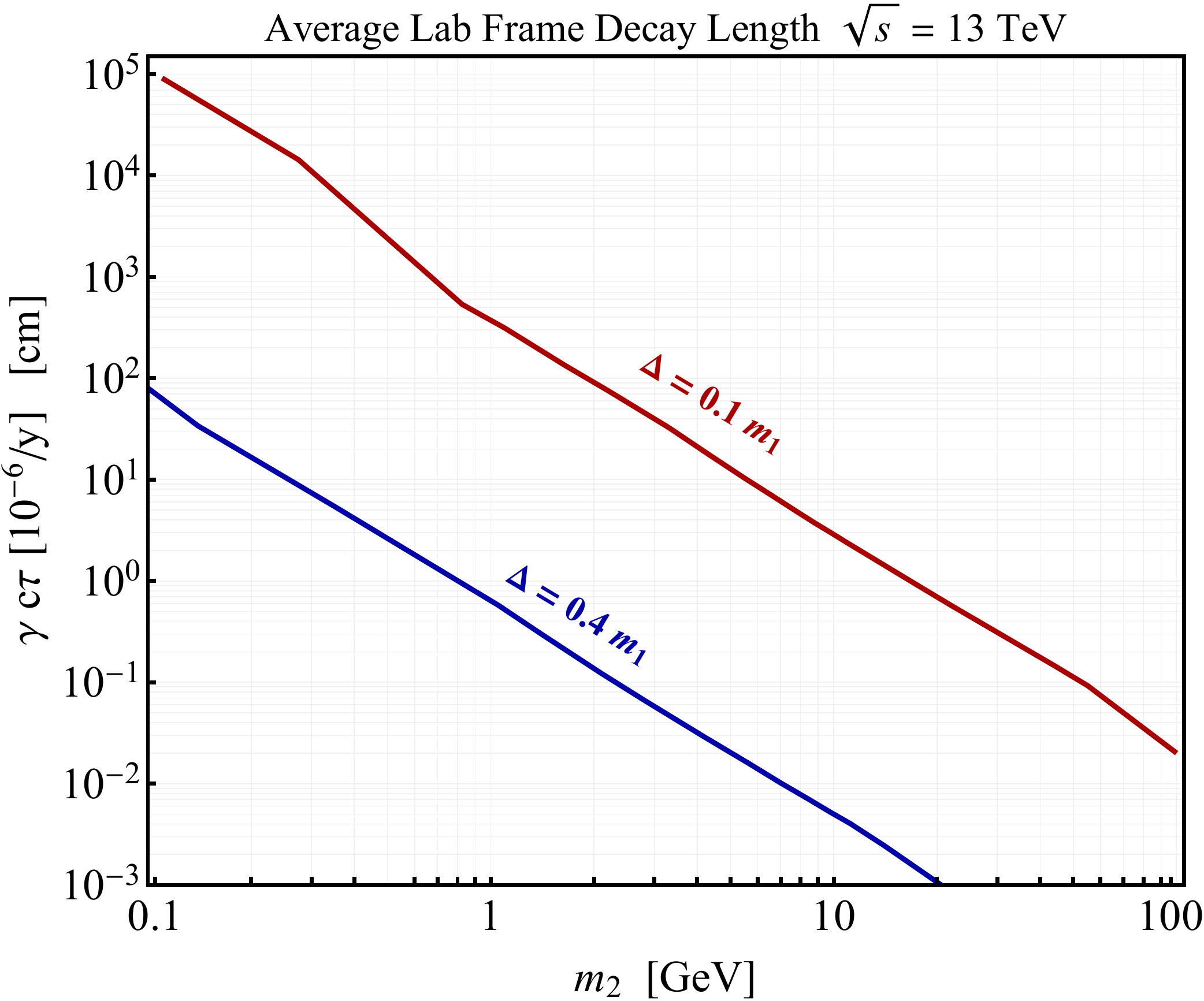}  
\caption{ Lab frame decay lengths in the fermion iDM dark photon model for boosted $\chi_2$ de-exictation via $\chi_2 \to \chi_1 \mu^+ \mu^-$, which is mediated by an off-shell $\apr$. Events are shown at $\sqrt{s}=13$ TeV and requiring a leading jet $p_{\rm T}>120~\GeV$. The results are normalized to $y = 10^{-6}$ for simple comparison with Figs.~\ref{fig:mainplot-fermion} -- \ref{fig:mainplot-scalar}. }
\label{fig:displacementPlot}
\end{figure}

We  describe the LHC signatures of the dark photon iDM model introduced in Sec.~\ref{sec:models}. We focus on the regime of few-GeV iDM masses, where existing constraints on thermal DM are relatively weak and for which dedicated collider searches must be developed to tag the SM states from $\mathrm{DM}^*$ decay.  In our model, the excited state $\mathrm{DM}^*$ decays via  ${\apr}^{(*)}$ to $\mathrm{DM}+\bar ff$ with the latter being SM fermions. Since $\Gamma_{\mathrm{DM}^*}\sim \Delta^5/m_{\apr}^4$ (see Appendix \ref{sec:appendixB-annihilation-rate}), $\mathrm{DM}^*$ is long-lived on collider scales for GeV-scale masses and moderate mass splittings ($\Delta/m_1 \sim0.01-0.1$), giving rise to decays within the LHC detectors at a displaced vertex. The signature is striking, but the leptons are typically both collimated and soft, motivating dedicated collider searches. We show a representative decay length distribution in Fig.~\ref{fig:displacementPlot}. A part of the parameter space was explored in Ref.~\cite{Bai:2011jg} in the context of iDM for fully hadronic $\chi_2\rightarrow \chi_1$ decays over a range of masses at fixed splitting, with a focus on contact operators with couplings to quarks\footnote{From this point on, we use $\chi_2$ and $\chi_1$ as placeholders for DM excited and ground states, respectively, regardless of whether DM is a scalar or fermion.}.  

In our representative model,  $A'$ couples to the quarks and leptons via hypercharge kinetic mixing. We  focus on the case where $m_{\apr} > m_2 + m_1$; in this scenario,  the $A'$ is produced on-shell in association with a jet\footnote{We also include DM production via $j+(Z\rightarrow\chi_2\chi_1)$, although this is subdominant for the parameters we study.} and the $A'$ decays to $\chi_1\chi_2$. The $\chi_2$ subsequently decays via an off-shell $A'$ to $\chi_1$ plus SM fermions. In particular, we focus on the decays to leptons, giving rise to a {\it displaced dilepton vertex} in association with a hard jet and missing momentum. In particular, we consider the reaction
\be
pp &\rightarrow& j + A' \rightarrow j \chi_2 \chi_1\\
&\rightarrow& j  \ell^+ \ell^- \chi_1 \chi_1.
\ee
The location of the displaced vertex determines the sensitivity of LHC searches to the $A'$ model. If $\chi_2$ escapes the detector before decaying, the signature reverts to that probed by  conventional monojet searches. If $\chi_2$ decays at distances $\lesssim1$ mm, then large (quasi-)prompt SM electroweak, top, and QCD backgrounds become important. We find the best sensitivity is in between these two regimes.

\begin{figure}[t] 
 \vspace{0.cm}
 ~~\includegraphics[width=8.5cm]{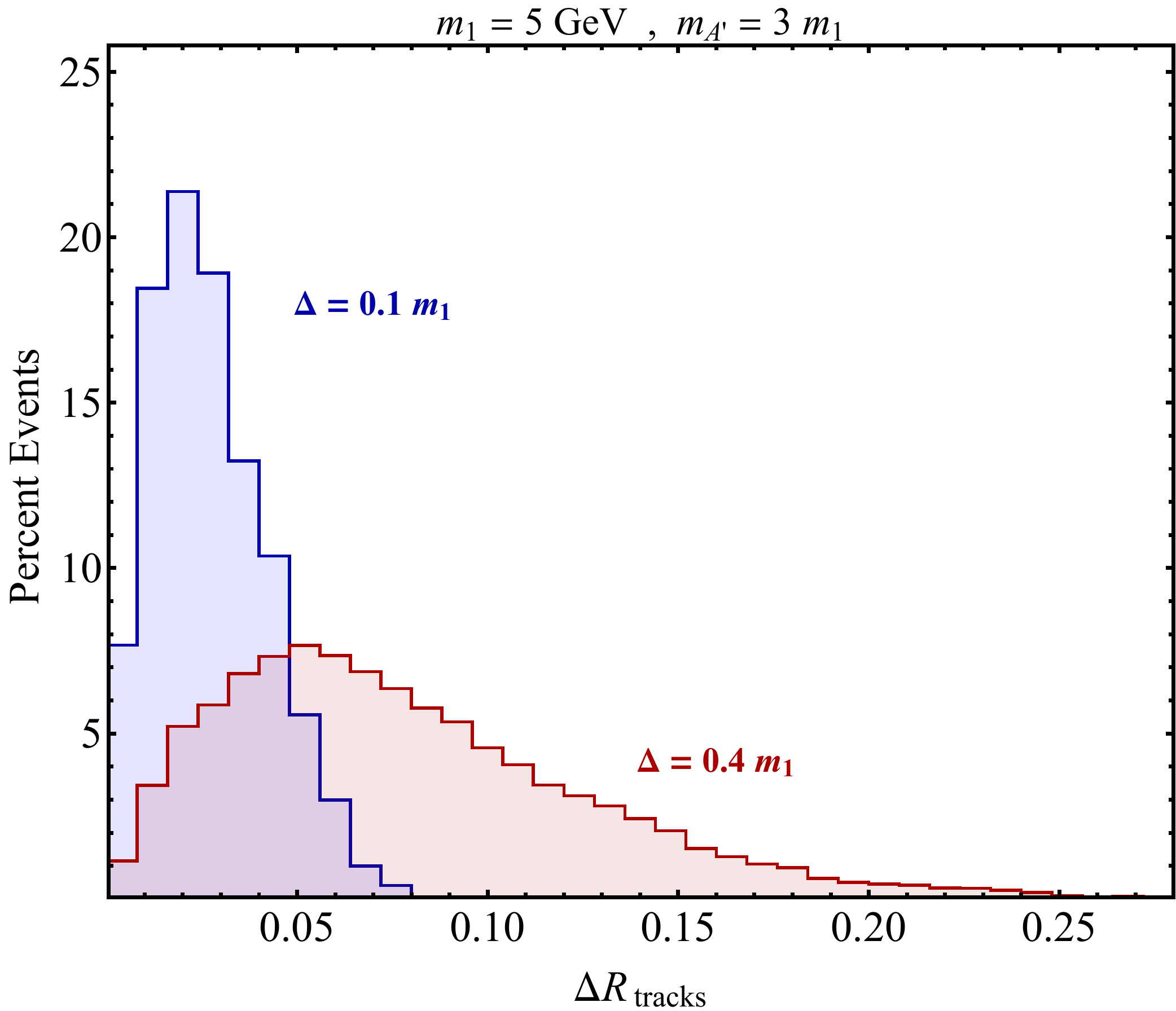}    
\caption{$\Delta R$ spectrum between the displaced dilepton tracks in the dark photon model. We consider $\Delta=0.1m_1$ and $\Delta=0.4m_1$, assuming $m_1=5~\GeV$ and $m_A=15~\GeV$.}
\label{fig:darkphotondeltaRtracks}
\end{figure}
\begin{figure*}[t] 
 \hspace{-0.5cm}
 ~~\includegraphics[width=8.65cm]{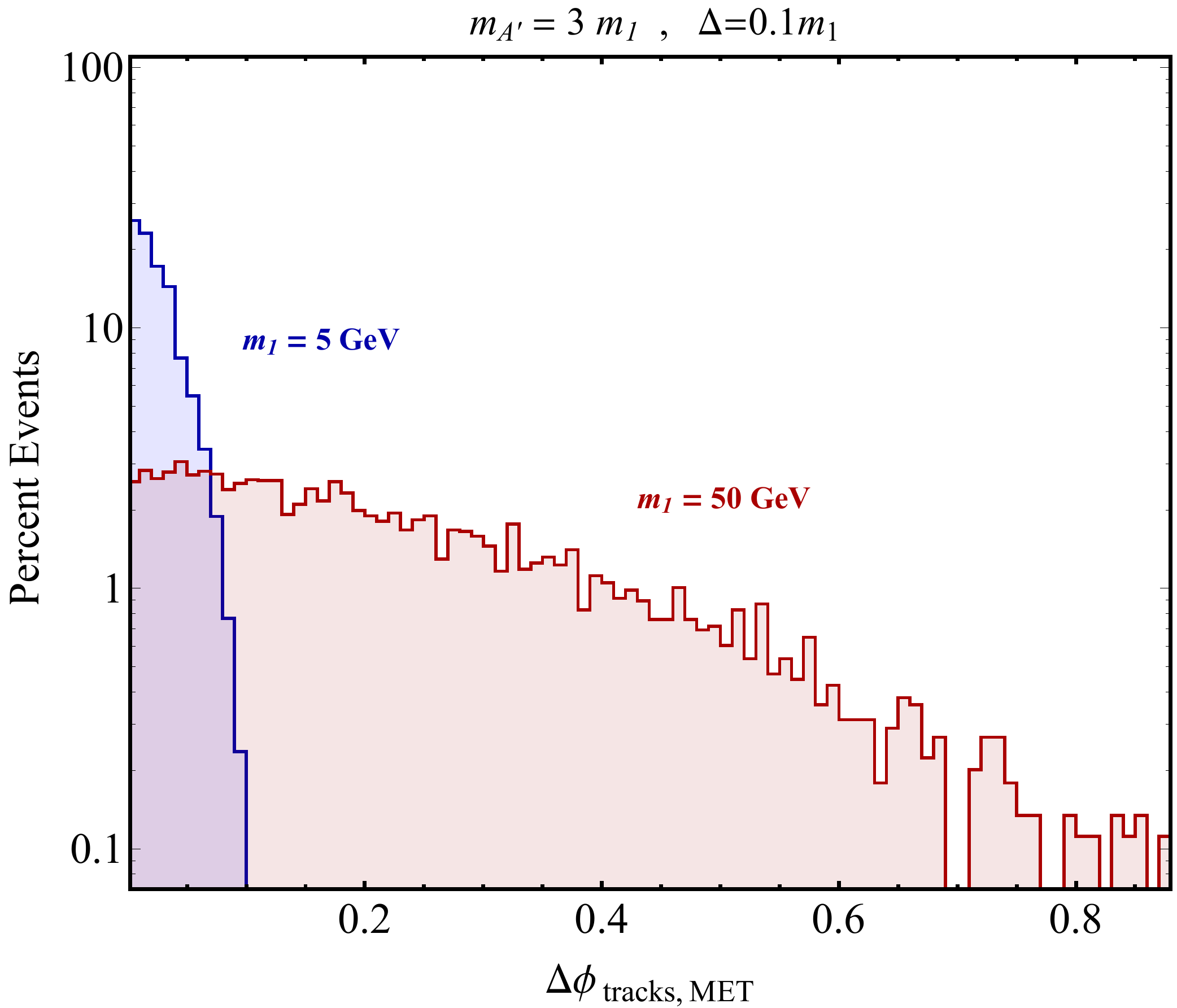}  
  ~~\includegraphics[width=8.65cm]{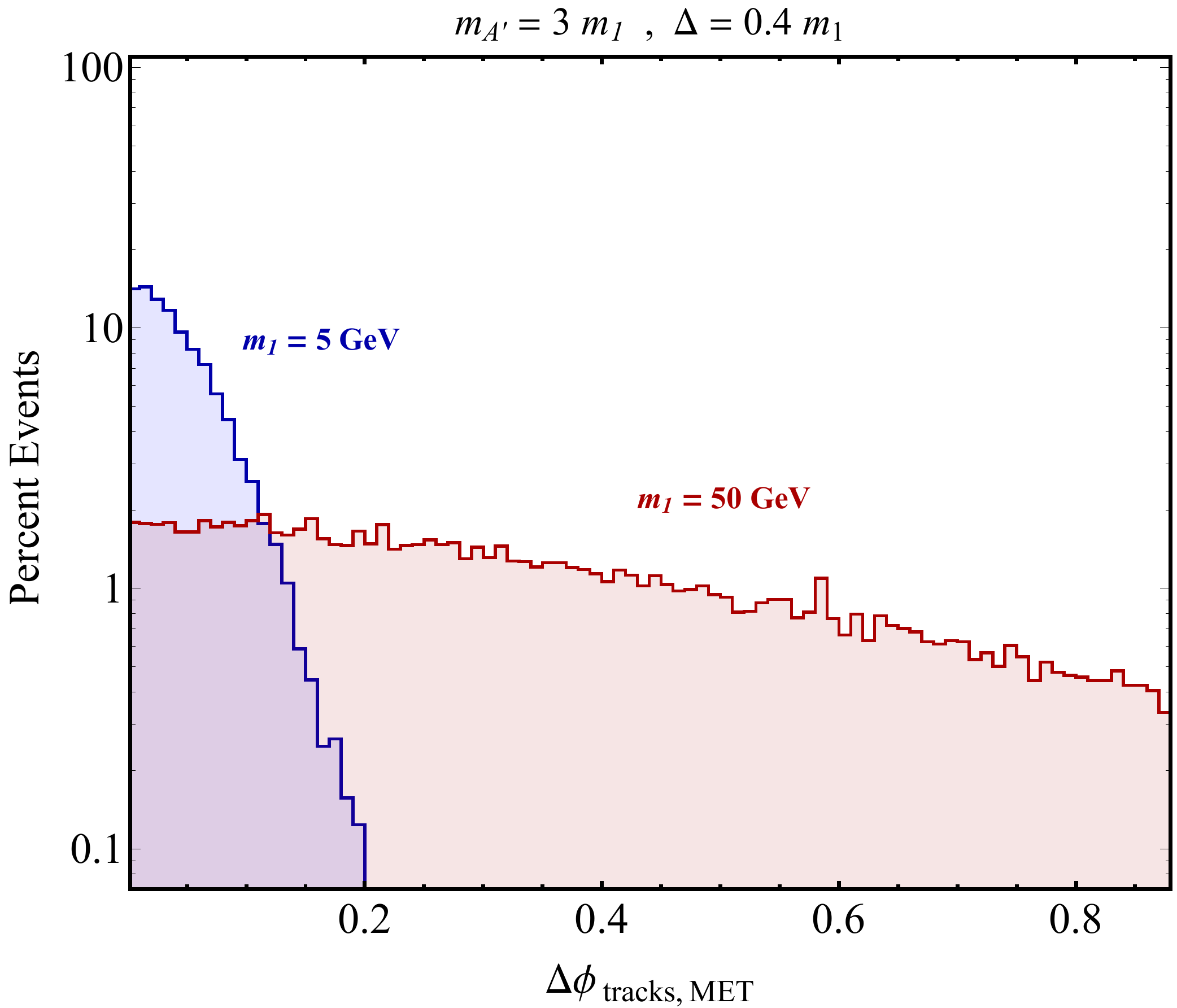}  
\caption{Distributions of $|\Delta\phi|$ between the momentum of the lepton jet and the $\slashed{E}_T$ of the event in the dark photon model. We consider $\Delta=0.1m_1$ and $\Delta=0.4m_1$, assuming $m_1=5~\GeV$ and $m_{\apr}=15~\GeV$.}
\label{fig:darkphotondeltaphi}
\end{figure*}

The signal also features distinctive kinematics for the dilepton pair. For $m_{\apr}\sim m_\chi\sim$ GeV, the $A'$ is highly boosted when recoiling off the high-$p_{\rm T}$ jet. As a result, the decay products are highly collimated in the decays $A'\rightarrow \chi_2\chi_1\rightarrow \ell^+\ell^-\chi_1\chi_1$. In particular, the leptons are well within $\Delta R\equiv\sqrt{\Delta \eta^2+\Delta \phi^2}<0.4$ of one another (see Fig.~\ref{fig:darkphotondeltaRtracks}  for a representative benchmark point), giving rise to a {\it lepton jet} signature (see Refs.~\cite{Strassler:2006im,Strassler:2006ri,Han:2007ae,Gopalakrishna:2008dv,ArkaniHamed:2008qp,Baumgart:2009tn,Cheung:2009su,Falkowski:2010cm,Izaguirre:2015pga,Buschmann:2015awa} for examples of lepton jets in other contexts). Similarly, the lepton jet will be collimated with the $\slashed{E}_{\rm T}$ of the event, as illustrated in Fig.~\ref{fig:darkphotondeltaphi}. In our analysis, we focus on the decays to muons since the backgrounds are smaller for this final state due to a lower fake rate. All of the above features of the signal combine to give a striking final state at the LHC, as we show below.

Ê

The backgrounds for displaced vertex searches are small, but difficult to estimate due to contributions from rare heavy-flavor decays as well as random track crossings in the detector. Monte Carlo (MC) tools do not always model the backgrounds well, which is why the experimental collaborations typically perform data-driven background estimates for long-lived particle searches. While we lack the tools and the data to do such an estimate, in what follows we discuss in turn the most important SM processes that can mimic the signal and perform rudimentary estimates of the importance for each, arguing that our proposed final state could well result in a virtually background-free search. This conclusion is also supported by the very low backgrounds observed  by ATLAS for soft displaced dilepton vertices \cite{Aad:2015rba};  the  trigger and event selections for this analysis require at least one very high $p_{\rm T}$ lepton or trackless jet, however, and so the analysis in its current form is not sensitive to dark photon iDM signatures. The background estimates we provide must, of course, be verified by detailed study by the experimental collaborations. 

In our analysis, we perform all signal and background calculations in this section with \texttt{Madgraph 5} \cite{Alwall:2011uj}, with subsequent showering and hadronization in \texttt{Pythia 6} \cite{Sjostrand:2006za}, and jet clustering with \texttt{Fastjet 3} \cite{Cacciari:2011ma}.  Moreover, we use \texttt{Feynrules} to define the two representative models in our study \cite{Alloul:2013bka}.\\

\noindent {\bf Backgrounds:}

\begin{figure*}[t] 
 \hspace{-0.5cm}
 ~~\includegraphics[width=8.5cm]{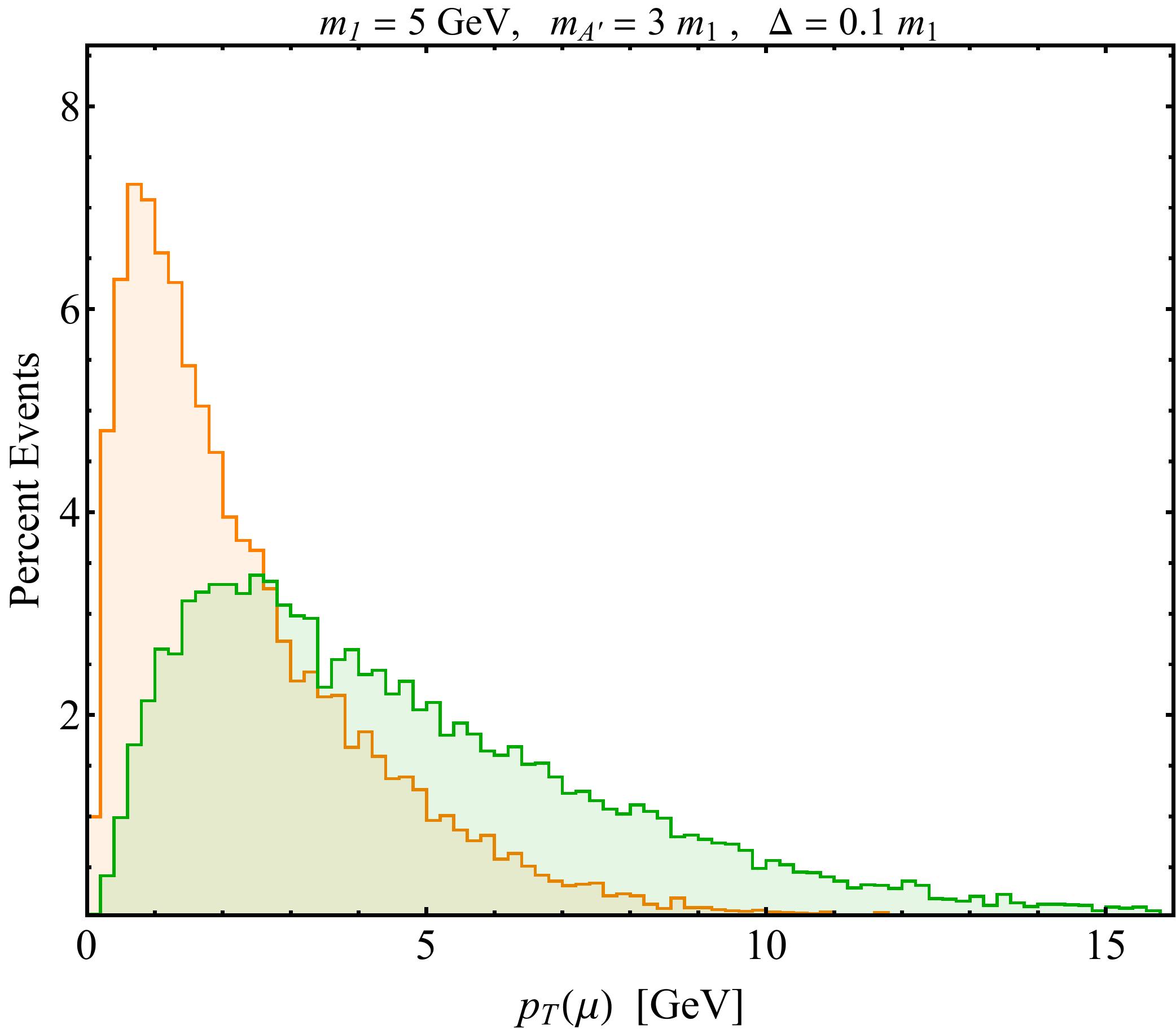}  
  ~~\includegraphics[width=8.5cm]{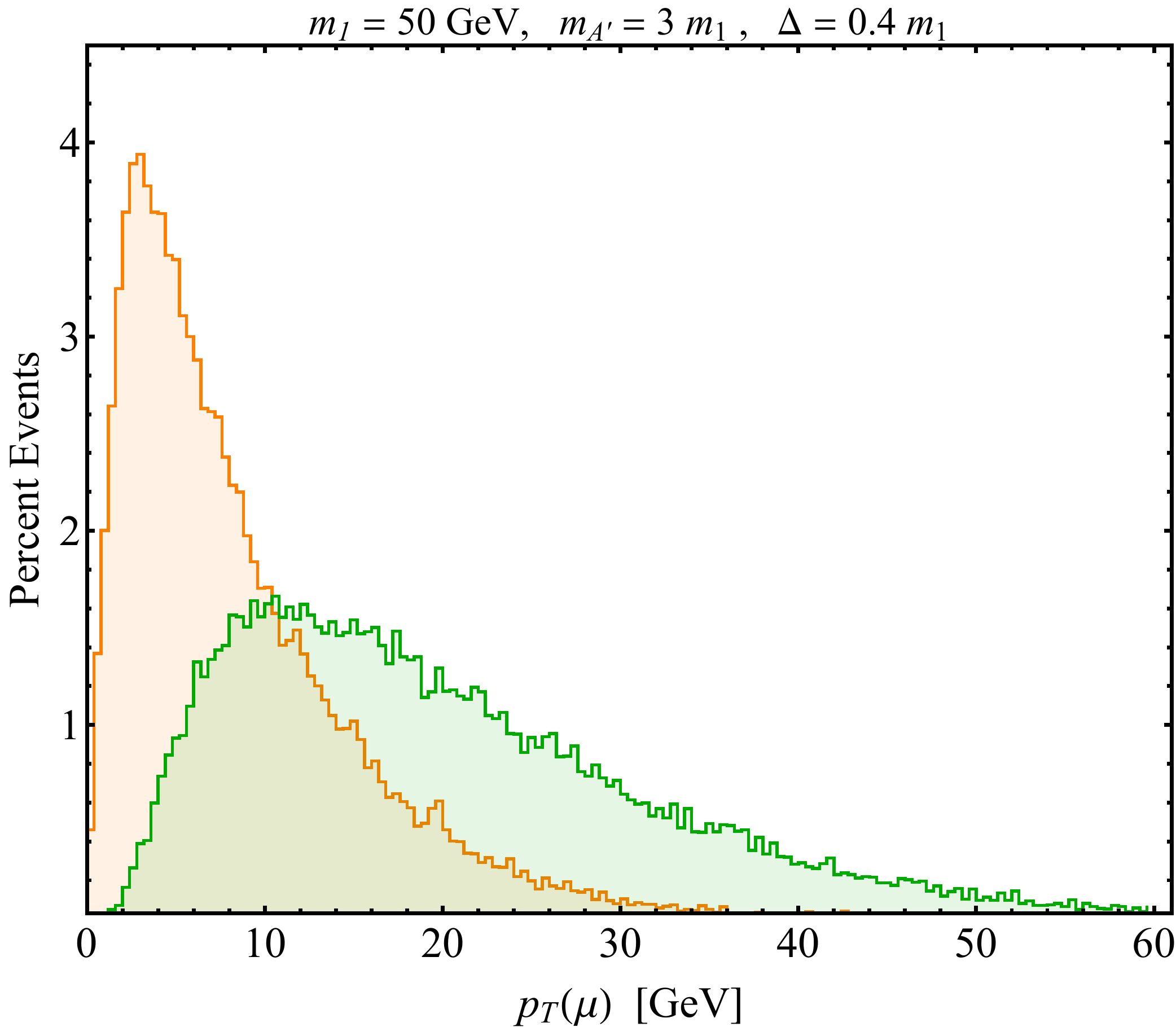}  
\caption{$p_T$ distributions for the leading and subleading muons in $\chi_2\to \chi_1 \mu^+ \mu^-$ decays at the LHC at $\sqrt{s}=13$ TeV in
the $\apr$-mediated scenario for representative masses and splittings. }
\label{fig:darkphotonmuonpT}
\end{figure*}

\begin{itemize}
\item {\it Real photon conversion}.
Non-prompt leptons can arise from the conversion of real photons via collisions with material or gas in the detector. A photon traversing the inner detector (ID) inside ATLAS or CMS will go through one radiation length of material, so an $\mathcal{O}(1)$-fraction of the photons convert into an $e^+e^-$ pair. Moreover, the production cross section for $pp\rightarrow j\gamma(Z\rightarrow\nu\bar\nu)$ at $\sqrt{s}=13~$ TeV is $\approx 100 $ fb, after requiring a jet with $p_{\rm T} > 120~\GeV$ and a sufficiently energetic, isolated photon. Nevertheless, there are several  handles to significantly reduce this background. First, events where the photon carries some of the momentum away can be rejected by applying stringent isolation requirements on the leptons. Second, leptons originating from material-dense parts of the detector and/or consistent with photon conversion can be vetoed.  Third, the invariant mass distribution of the leptons from photon conversions will peak at zero mass as they originate from an on-shell photon. Finally, the photon conversion probability to muons is suppressed relative to electrons by $\frac{m^2_e}{m^2_{\mu}}\approx 10^{-5}$. All of these considerations combine to allow an estimate of negligible background from photon conversions.

\item {\it QCD}. Displaced tracks could originate from QCD-initiated jets, particularly those giving rise to long-lived $B$ or $K$ hadrons which in turn decay to $\pi$ and/or $\mu$. Estimating this background from first principles is not feasible due to the dependence on hadronization effects and the challenge of estimating muon mis-identification rates; nonetheless, we  determine an approximate upper bound on the probability for a QCD-initiated event to give a hard leading jet with $p_{\rm T} > 120~\GeV$, and {\it two} muon-tracks with $p_T > 5~\GeV$ appearing at a displaced vertex (both muon tracks have transverse impact parameter $|d_0|$ between 1 mm and 30 cm, and the point of closest approach of the tracks is $<1$ mm). We estimate this probability to be $<10^{-7}$, which bounds the QCD cross section to be $ \lesssim 100$ fb. This requirement is {\it before} requiring significant missing energy from hadrons in the event, and before requiring that the missing energy be near the muon-tracks or any other kinematic features characteristic of signal.

\item {\it Pile-up}. In the QCD estimate, we assumed that the jet and the muons from long-lived hadron decays originate from the same primary vertex. In upcoming running of the LHC, the increased luminosity comes with the price of a large number of primary vertices per bunch crossing due to pile-up. Therefore, it is possible that the soft, displaced muons could originate from a different primary vertex. Since the signal muons are highly collimated, however, they point in the same direction as the long-lived particle, which passes through the primary vertex with the high-$p_{\rm T}$ jet. Therefore, even though each muon has a high impact parameter, applying a selection requirement that the dimuon momentum approximately point back to the primary vertex can be used to suppress long-lived hadronic backgrounds from other primary vertices.

\item {\it Jet + di-tau}. 
The cross section for a high-$p_{\rm T}$ jet, along with two $\tau$ leptons within $\Delta R<0.01$ of one another, at the LHC at $\sqrt{s}=13~\TeV$  is $\approx 10$ fb. Accounting for the requirement that both taus decay to a muon further reduces this rate to $\sim 10^{-1}$ fb. In addition, for this background component to mimic the signal, both taus need to decay within  $\sim100$ $\mu$m of each other. And finally, we note that since each muon-track originates from a different $\tau$ parent, the $m_{\mu\mu}$ distribution will be distinct from the signal where both tracks originate from the $\chi_2$.

\item {\it Jets + $V\rightarrow \slashed{E}_T$}. A potential background may originate from the reaction $pp\rightarrow \mathrm{jets}+V$, with $V$ either a $Z$ or a $W$ boson decaying to give missing energy. For this background to contaminate the signal region, one would need the two tracks to originate from the jets. Through a reasoning analogous to that used above for the QCD background, this background component should be in the range of less than $\sim 0.01$ fb, and so relatively negligible for our analyses.

\item {\it Backgrounds from fake missing energy}. Typically, experimental analyses require a minimum separation between missing energy and other objects in the event to suppress fake missing energy from calorimeter or momentum mis-measurement. By contrast, our signal is collimated with the missing energy, and so fake missing energy is a potential concern. We exploit the fact that the muons from signal decays are relatively soft (typically, the summed muon $p_{\rm T}/\slashed{E}_{\rm T}$ is $\lesssim0.2$; see Fig.~\ref{fig:darkphotonmuonpT}), and so fake missing momentum is not expected to be important.

\end{itemize} 

\noindent {\bf Signal region:} The above considerations motivate the following selections for the signal region:

\begin{itemize}
\item Trigger on a monojet + $\cancel{E}_{\rm T}$. For Run 1, for instance, CMS used a $\slashed{H}_T>120~\GeV$ trigger \cite{Khachatryan:2014rra}, where $\slashed{H}_T$ is the missing momentum as computed in all subsystems excluding the muon system. We assume such a trigger for our sensitivity estimates for $\sqrt{s}=13~\TeV$, although note that the exact values for Run 2 could be higher. The additional use of the soft leptons could help keep trigger thresholds low; for example, ATLAS has an analysis which has requirements as low as $p_{\rm T}>6~\GeV$ for muons at trigger level \cite{Aad:2014yea}. Nevertheless,  we also checked that with a trigger of $p_{\mathrm{T}j}>200$ GeV and $\cancel{H}_{\rm T}>200$ GeV, the signal sensitivity is degraded by approximately a factor of two in rate;
\item One leading jet with $p_T > 120~\GeV$ and allow only one extra jet with $p_{\rm T}>30$ GeV; the leading jet and $\cancel{H}_{\rm T}$ should be back-to-back;
\item One displaced muon jet, $\mu J$, consisting of at least two muons with $|d_0|$ between 1 mm and 30 cm, and whose tracks cross within 1 mm; the two muon tracks each have $p_T > 5~\GeV$;
\item $\slashed{H}_T > 120~\GeV$;
\item $|\Delta\phi(\slashed{E}_T,\mu J)| < 0.4$. 
\end{itemize}

We show our projections for the LHC sensitivity to this topology at $\sqrt{s}=13~\TeV$, assuming $\mathcal{L}=300$ fb$^{-1}$ of integrated luminosity in Fig.~\ref{fig:mainplot-fermion} through Fig.~\ref{fig:mainplot-scalar} for fermionic and scalar DM. As motivated above, we assume a mostly background-free signal region, and therefore plot sensitivities for 10 signal events. The unique kinematics of the signals we study could allow the LHC to probe the very couplings responsible for establishing the DM abundance in the early universe through thermal freeze-out.

\subsubsection*{$B$ factories}

$B$-factories have the potential to make significant progress in the exploration of DM with inelastic interactions for DM masses within kinematic reach (below a few GeV). This is due to the very high luminosity and clean environment of an $e^+e^-$ collider.

There are two potential avenues to pursue at a B-factory. One possibility is to search for direct production of the ground and excited states with subsequent decay of the excited state to (displaced) tracks. It is unclear, however, whether such displaced tracks are sufficiently energetic or well-reconstructed to pass the trigger in analyses such as Ref.~\cite{Lees:2015rxq}. Alternatively, one can trigger on and reconstruct a visible SM state $Y$ produced in association with DM ($e^+e^-\rightarrow \mathrm{DM}+\mathrm{DM}^{*} + Y$). In particular, the BaBar experiment instrumented a monophoton trigger for $\approx 60$ fb$^{-1}$ of the total dataset. In our study, we conservatively  consider only the latter scenario as a trigger for our proposals, although both possibilities should be investigated by $B$-factories.  We base the following  results on the analysis from Ref.~\cite{Aubert:2008as}, which used a photon trigger with threshold $E_\gamma > 2~\GeV$.\\

\noindent {\bf Monophoton + missing energy}: The analysis from Ref.~\cite{Aubert:2008as} performed a search for (untagged) decays of $\Upsilon(3S)\rightarrow A^0+\gamma$ ($\mathcal{L}\approx 25 $ fb$^{-1}$), where $A^{0}$ is an invisibly-decaying pseudoscalar, with a stringent veto on additional activity in the detector. The dark photon in our model is produced through the reaction $e^{+}e^{-}\rightarrow \gamma + A'$, with subsequent decay $A'\rightarrow \chi_2 \chi_1$. Although our model produces visible states in $\chi_2$ decays, the kinematics of the dark photon signal can still populate the BaBar signal region, which consisted of a bump search in the missing mass variable $m^2_X = m^2_{\Upsilon(3S)} - 2 E_{\gamma,\rm{CM}}m_{\Upsilon(3S)}$.  For our signal to appear in this search, the $\chi_2$ has to decay either outside the detector or into soft final states that fall below BaBar's thresholds (we use the thresholds listed in Ref.~\cite{Aubert:2001tu}). 

A complication of this analysis is that, for $m_{\apr}<1~\GeV$, the signal could appear in the BaBar control region, in which case the signal mimics the kinematics of the irreducible $\gamma\gamma$ background. In this mass regime, we set a conservative bound by assuming that the signal represents all of the events in the control region. 

The results of the BaBar monophoton + missing energy recast are shown in Fig.~\ref{fig:mainplot-fermion} to \ref{fig:mainplot-scalar}. We also provide a projection for Belle II assuming $50\,\,\mathrm{ab}^{-1}$ of luminosity with an instrumented monophoton trigger \cite{Essig:2013vha}.  For a more extensive discussion of the details of this analysis, see Ref.~\cite{Izaguirre:2013uxa} and Ref.~\cite{Essig:2013vha}.\\

\noindent {\bf Monophoton + displaced tracks + missing energy:} A potentially more striking signature of iDM at $B$-factories could be uncovered by a re-analysis of BaBar data. In particular, the reaction $e^{+}e^{-}\rightarrow \gamma + A'$ can give rise to displaced tracks and missing momentum in the final state. As before, we assume use of the monophoton trigger, and offline selection of two displaced leptons with $p > 100~\MeV$ and transverse impact parameter $|d_0|$ between 1 cm and 50 cm, as in Ref.~\cite{Lees:2015rxq}. Based on Ref.~\cite{Allmendinger:2012ch}, we use a lepton reconstruction efficiency of 50\% in our estimates.

We expect backgrounds from resonances (arising from the decay of a hadron, or through radiative return) in this channel to be low, particularly after requiring significant $\slashed{E}$ and removing dilepton pairs consistent with hadronic resonances. Studies of backgrounds in related searches for $B^0\rightarrow J/\psi\,\gamma$ \cite{Aubert:2004xd} and radiative decays of $\Upsilon\rightarrow \gamma (A^0\rightarrow\mu^+\mu^-)$ \cite{Aubert:2009cp} (where $A^0$ is a light exotic scalar) suggest that this may indeed be the case for our proposed channel. Additionally, from Ref.~\cite{Lees:2012cj}, another potential background is that from $\gamma \pi^+ \pi^-$, and $\gamma \gamma$ with one of the $\gamma$ converting to a $\ell^+\ell^-$ pair. However, the former can be reduced with the requirement that the tracks originate from a high impact parameter vertex, and the both the former and the latter could be reduced through a combination of a missing mass cut and a cut on the invariant mass of the tracks.\\

\noindent {\bf Results:} The above proposed searches at BaBar prove complementary to the searches at the LHC that we advocate. In particular, we find they have the potential to cover thermal-relic territory for the $\mathcal{O}(10)$\% fractional mass splittings that are the focus of our analysis. Figs.~\ref{fig:mainplot-fermion}  -- \ref{fig:mainplot-scalar} illustrate the potentially powerful reach that BaBar could achieve with a dedicated monophoton + displaced tracks search. Additional improvements could be achieved by future $B$-factories \cite{Essig:2013vha} depending on whether or not they are instrumented with a monophoton trigger, especially outside of the control region where the sensitivity scales as $\sqrt{\mathcal{L}}$; therefore, our analysis provides further motivation for the development of a monophoton trigger for Belle II.

\subsection{Magnetic Dipole Interaction}

\subsubsection*{LHC}

The second simplified model we consider is dark matter coupled inelastically via a magnetic dipole moment (see Sec.~\ref{sec:models}). In this scenario, the excited DM state $\chi_2$ decays via $\chi_2\rightarrow\chi_1+\gamma$. We are interested specifically in the $m_\chi\sim100$ MeV-100 GeV, $\mathcal{O}(10\%)$ splitting inelastic limit considered earlier. As before, the decay products of $\chi_2\rightarrow \chi_1 + \gamma$ are typically too soft to serve as the main trigger objects, and so we rely on the associated production of a high-$p_{\rm T}$ jet. Thus, we predict a $pp\rightarrow j+\slashed{E}_{\rm T}+\gamma$ signature. Existing work has studied the scenario with a hard photon originating from larger splittings between DM states in both the prompt and long-lived limits \cite{Weiner:2012cb,Primulando:2015lfa}.

 There are two principal distinctions between the dipole scenario and the dark photon considered earlier. The dipole is a dimension-5 operator, and so the decay width of $\chi_2$ through the dipole $\mu_\gamma$ in the limit of small splittings $\Delta$ goes like $\Gamma\sim \mu_\gamma^2\Delta^3$ (see Appendix \ref{sec:appendixB-annihilation-rate}); by contrast, decays through an off-shell dark photon scale like $\Delta^5/m_{A'}^4$ and is suppressed by 3-body phase space. As a result, the decays are prompt over a wide range of the dipole parameter space, and consequently the backgrounds are significantly larger than in the displaced muon jet analysis. Furthermore, it is more challenging to reconstruct soft photons than soft muons, with the photon reconstruction efficiency $>$ 0.5 only above $E_{\rm T}=15$ GeV (see, for example, Ref.~\cite{ATLAS-CONF-2012-123}). Thus, the sensitivity of a dedicated search for the existence and kinematics of the soft photon is lower than for the dimuons. Nevertheless, we find that dedicated monojet + photon + missing energy searches can be competitive with existing limits and cover interesting, unexplored parts of the parameter space, particularly for $B$-factory probes.
 
We consider two cases:~in the first, $\chi_2$ decays on lengths shorter than the pointing resolution of the electromagnetic calorimeter (ECAL), namely $\lesssim25$ mm for $E_{\mathrm{T}\gamma}\approx20$ GeV \cite{Aad:2014gfa}. This region of lifetimes can be probed by dressing the canonical monojet searches with a prompt, soft photon requirement. Such searches have significant electroweak backgrounds, but our proposed searches are competitive with existing constraints and can  have some sensitivity over new parameter space due to the distinctive signal kinematics. Second, for small dipole moments, the decay occurs on macroscopic scales and the photon from $\chi_2$ decay does not point back to the primary vertex. Such non-pointing photon signatures have been considered in SUSY searches \cite{ATLAS-CONF-2012-123} for pairs of photons. For our signal, $\chi_2$ is highly boosted if the photon is hard enough to be seen in the detector, and so it still points approximately in the direction of the primary vertex; consequently, when the photon decays before the ECAL, its pointing displacement is typically $\lesssim$ cm, and so cannot be effectively distinguished from prompt decays. Thus, we focus on prompt searches here.

We begin by examining the relevant  backgrounds. In our analysis, we use the same MC tools as for the dark photon model described above. We re-normalize backgrounds  using an average next-to-leading order $K$-factor calculated using \texttt{MCFM} \cite{Campbell:2000bg,Campbell:2010ff}; as we show below, the signal over most parts of parameter space is produced via $(Z\rightarrow\chi_2\chi_1)+j$, and so we apply the $Z$+jets $K$-factor to signal as well.\\

\noindent {\bf Backgrounds:}

\begin{itemize}

\item {\it Jets + $\gamma+\ell+\nu$}.
This dominant background comes from $W$+jets, where the lepton undergoes bremsstrahlung and emits a photon. While lepton vetoes can be used to suppress this background, a hard photon and soft lepton occur in a substantial fraction of events. The resulting photon is correlated with the direction of the lepton, and for boosted $W$ bosons, this results in a soft photon aligned with $\slashed{E}_{\rm T}$. This provides an irreducible background mimicking our signal. We assume that photons produced by collisions of the lepton with gas and detector material can be suppressed through careful event selection, and so the dominant background comes from prompt photons. Applying the lepton veto from the ATLAS analysis \cite{Aad:2015zva} and a sample of preselection requirements ($p_{\mathrm{T}j}>120$ GeV, $\slashed{E}_{\rm T}>100$ GeV,  an isolated $E_{\mathrm{T}\gamma}>15$ GeV), and $|\Delta\phi(\slashed{E}_{\rm T},\gamma)|<1.5$, this background has a cross section of approximately 250 fb  at $\sqrt{s}=13$ TeV.

\item {\it Jet + $(Z\rightarrow \bar\nu\nu)$}. This is the dominant background for monojet searches at the LHC, and it produces photons through decays of pions and other hadrons. However, the photons are typically neither energetic nor isolated, and the presence of nearby hadronic activity can be used to greatly suppress this background. After requiring the same preselection cuts as for $W$+jets, the cross section is approximately 5 fb. The photon is typically aligned with the jet, and unlike for the signal and $W$ + jets background,  this can be used to provide significant discrimination. There may also be fakes where an additional jet fakes a photon, although we expect that they can be suppressed with sufficiently tight photon identification requirements so that the rate is subdominant to the $W$+jets background. 

\item {\it Jet + $\gamma+(Z\rightarrow \bar\nu \nu)$}.
This background is the same as the above but with the inclusion of a hard, prompt photon. Its cross section after pre-selection is approximately 50 fb, which is larger than the pure $Z$+jets rate. The $Z$+jets background has very similar kinematics to the $Z$+$\gamma$+jets, with the photon typically correlated with the direction of the jet.

\item {\it Fake missing energy}.
Because the photon is aligned with the missing energy, it is possible that QCD backgrounds fake both missing energy from jet mismeasurement as well as a photon from the same jet. We cannot simulate this background. However, the ratio of $E_{\mathrm{T}\gamma}/\slashed{E}_{\rm T}$ is typically small ($\lesssim20\%$) for signal and large for fake backgrounds, and so we expect that it can be greatly suppressed through $\slashed{E}_{\rm T}$ isolation requirements. We  expect that this background should be negligible, and can be suppressed by increasing the $p_{\rm T}$ and $\slashed{E}_{\rm T}$ thresholds if necessary.

\end{itemize}

While there are some fake backgrounds that we cannot simulate, the above estimates are sufficient for a qualitative estimate of the parameter space reach of a monojet + photon analysis.\\

\noindent {\bf Signal region:} We employ a monojet + missing energy trigger. At 8 TeV, the lowest threshold monojet + $\slashed{E}_{\rm T}$ trigger required monojet $p_{\rm T}>80~\GeV$  and $\slashed{E}_{\rm T}>105~\GeV$  \cite{Khachatryan:2014rra}, with the efficiency plateau somewhat higher. This threshold  may increase at higher energies; however, if the photon from $\chi_2$ decay is sufficiently hard, it may  be used to lower the monojet trigger threshold\footnote{In the  CMS high-level trigger at 8 TeV, photons as soft as $E_{\rm T}>23$ GeV were used to keep the diphoton trigger threshold low \cite{Trocino:2014jya}).}. Therefore, we perform an analysis with this trigger as the minimum threshold. We also briefly consider the effect of a higher threshold:~for example, if we consider instead a minimum threshold of $p_{\mathrm{T}j}>200$ GeV and $\slashed{E}_{\rm T}>200$ GeV, the results are nearly unchanged for $\Delta\lesssim0.1m_1$ because, in this regime, the $\chi_2$ needs a large boost for the photon to mass minimum reconstruction thresholds, driving the signal events to the large $p_{\rm T}$ and $\slashed{E}_{\rm T}$ regime. For larger $\Delta$, less boost is needed for signal events to pass the photon cuts, and we find a degradation of  $\lsim25\%$ for the prompt search. This does not qualitatively change our results.

\begin{figure*}[t] 
 ~~\includegraphics[width=8.5cm]{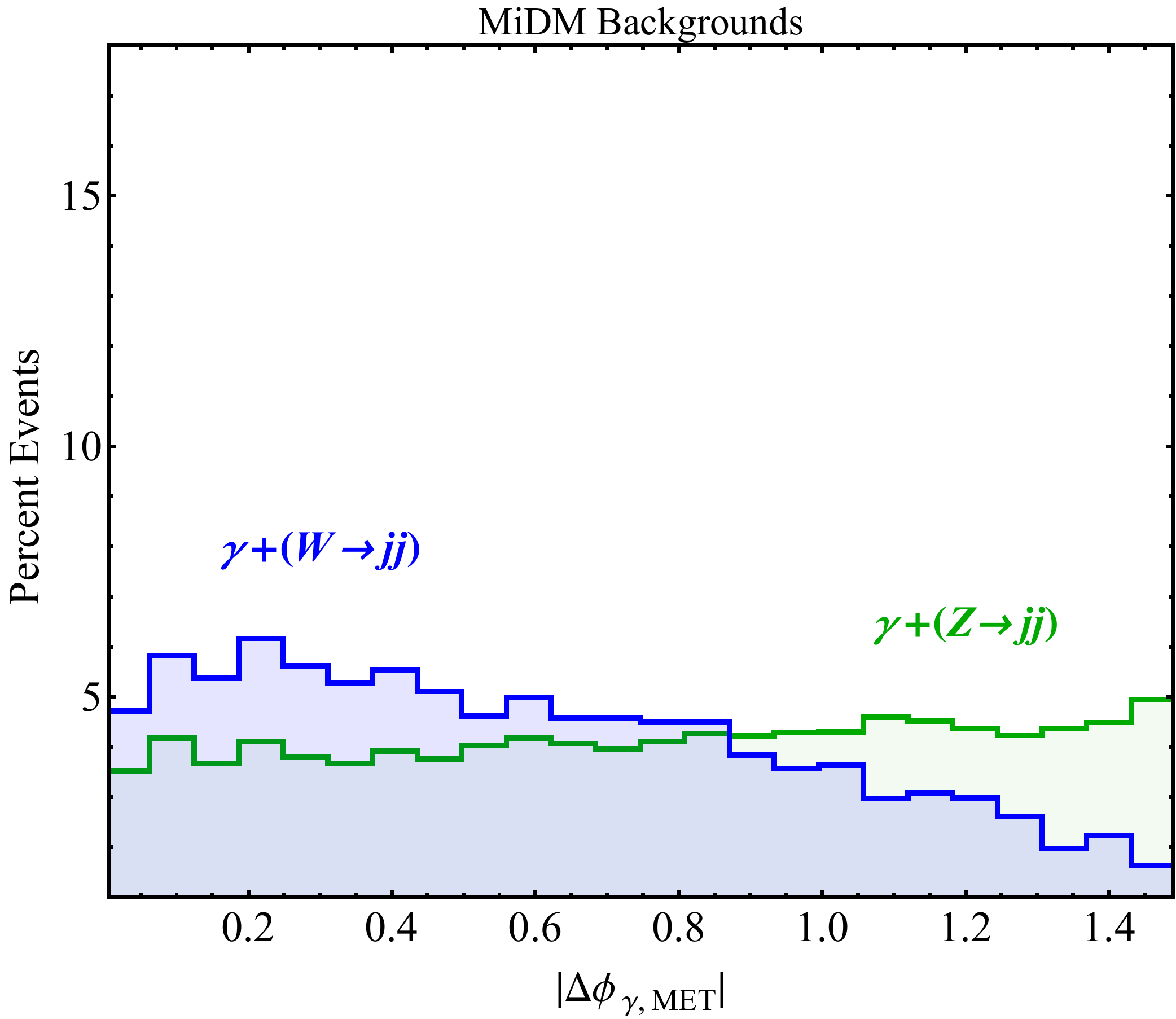} 
  ~~\includegraphics[width=8.5cm]{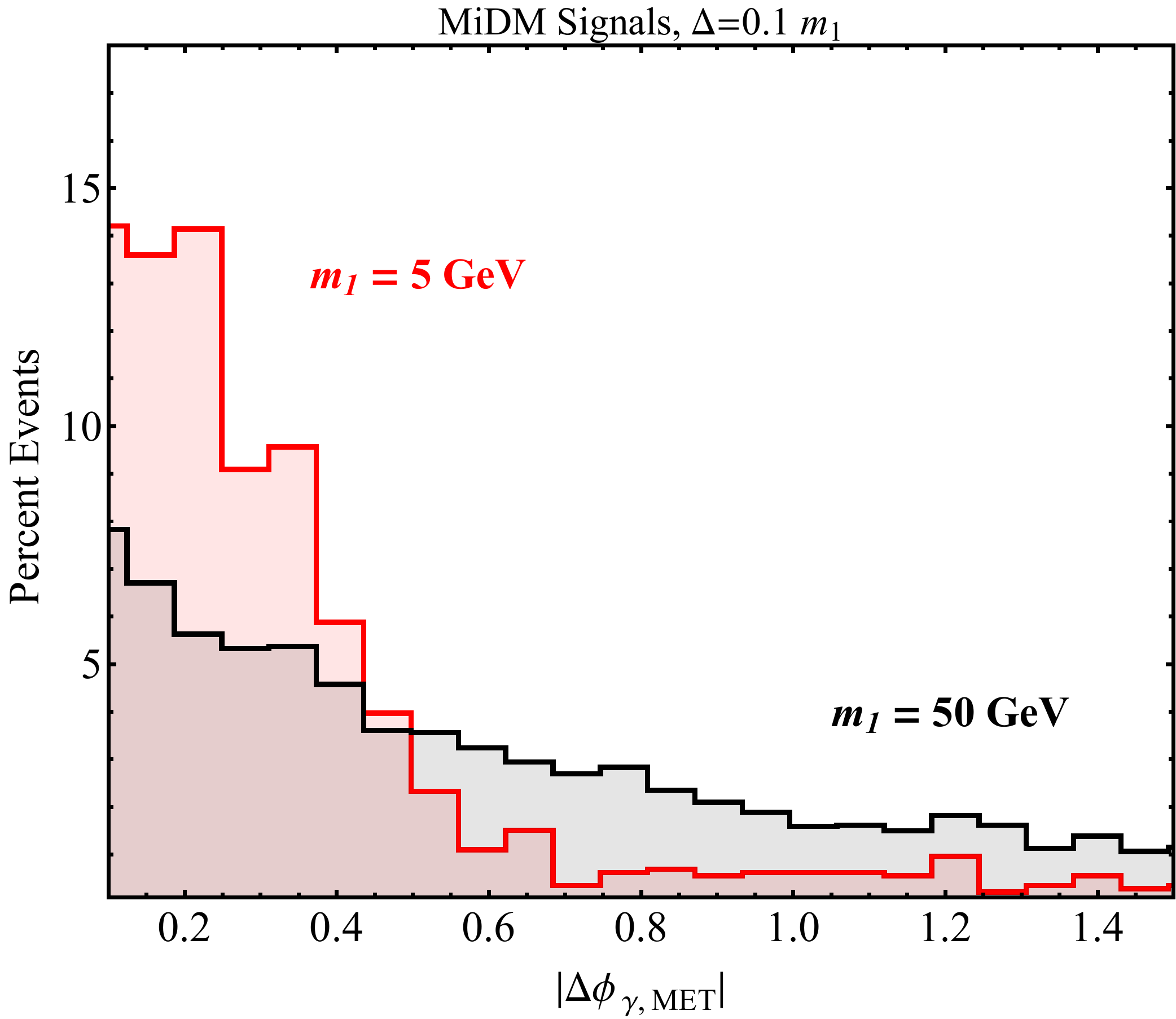}  
\caption{Distributions of $|\Delta\phi|$  between the momentum of the photon and the $\slashed{E}_T$ of the event for SM backgrounds (left) and signal in the dipole model (right). For signal, we consider $\Delta=0.1m_1$, $m_1=5$ GeV and 50 GeV.}
\label{fig:deltaPhiMiDM}
\end{figure*}

The backgrounds presented above limit the effectiveness of the prompt analysis and motivate the  signal region for our analysis, which is obtained by cutting on various kinematic observables. We scan over values of the cuts, and the least constraining values of each cut are listed below:
\begin{itemize}
\item One central hard jet ($|\eta|<2.5$,  $p_{\rm T}>120$ GeV) and allow only one extra jet with $p_{\rm T}>30$ GeV; the leading jet and $\cancel{E}_{\rm T}$ should be back-to-back;

\item Missing momentum $\slashed{E}_{\rm T}>105$ GeV;

\item One isolated\footnote{We require no more than 4 GeV of energy deposited within $\Delta R<0.4$ of the photon, not including the photon itself.} central photon (minimum $E_{\rm T} >15~\GeV$). We apply a flat 50\% identification efficiency for the photon, which is appropriate for $E_{\rm T}\approx15~\GeV$ \cite{ATLAS-CONF-2012-123};

\item Difference in azimuthal angle $|\Delta\phi(\slashed{E}_{\rm T},\gamma)| < 1$, as motivated by Fig.~\ref{fig:deltaPhiMiDM};

\item Transverse mass: \\
$m_{\rm T} \equiv \sqrt{2\slashed{E}_{\rm T}E_{\mathrm{T}\gamma}[1-\cos\Delta\phi(\slashed{E}_{\rm T},\gamma)]}<50$ GeV;

\item The long-lived particle decay must occur before reaching the ECAL and faster than 4 ns. We also impose a pointing requirement:~we calculate the distance $|\Delta z_\gamma|$ by using the photon direction to determine the point of closest approach to the primary vertex in the transverse plane, and then defining $|\Delta z_\gamma|$ to be the longitudinal distance between this point and the primary vertex \cite{Aad:2014gfa}. We require that $|\Delta z_\gamma|$ be within the detector resolution, and so satisfies $|\Delta z_\gamma|\lesssim15\,\,\mathrm{mm}\sqrt{50\,\,\mathrm{GeV}/E_{\mathrm{T}\gamma}}$ .
\end{itemize}

The transverse mass and $\Delta \phi$ observables are somewhat correlated, but we find performance is up to 25\% better with both included. For each model benchmark point, we optimize over simple cuts in each observable, and use this to determine the $2\sigma$ reach at 300 $\mathrm{fb}^{-1}$ assuming a systematic uncertainty of 10\% (and requiring a minimum number of three signal events). \\

\noindent {\bf Results:} We show our results in Fig.~\ref{fig:mainplot-dipole} under ``LHC prompt''. Over the range of splittings considered, our proposed search is competitive with existing bounds and can constrain open parameter space for MiDM. The cutoff of the sensitivity at $m_1\sim50$ GeV is not due to any kinematic effect, but rather due to a breakdown of the MiDM effective theory as we discuss below. Nevertheless, we expect the LHC to have reach for higher masses in a UV-complete theory.\\

\noindent {\bf Validity of the Dipole Effective Field Theory}:
The dipole operator is simply the lowest term in a series expansion generated by a UV complete theory with new mass states at some scale $\Lambda$. When the momentum flowing through the photon, $q^2$, becomes larger than the scale of the new mass states, then the new states can appear on-shell and invalidate the effective field theory (EFT). In the case of the dipole operator, the largest scale $\Lambda$ at which new states appear is $\Lambda\approx e/\mu_\gamma$, which corresponds to UV physics that saturates the perturbativity limit. Therefore, our analysis necessarily breaks down near $q^2 \sim \Lambda^2$, and processes with $q^2>\Lambda^2$ are not valid unless resolved by a form factor. 

To avoid sensitivity to an unphysical regime, we veto events where the dipole EFT description is invalid in calculating our results shown in Fig.~\ref{fig:mainplot-dipole} \cite{Busoni:2014sya,Aad:2015zva,Jacques:2015zha}. The valid MC events in our analysis often come from $(Z\rightarrow\chi_2\chi_1)+j$, which explains the cutoff in sensitivity due to EFT invalidity for $m_1>m_Z/2$ in Fig.~\ref{fig:mainplot-dipole}. A study of Dirac MiDM models in a particular UV completion \cite{Primulando:2015lfa} found that the EFT description was {\it conservative} for rates with $p_{\rm T}$ lower than the cutoff, since it neglects the typically large contribution to the cross section for $q^2 \sim \Lambda^2$ (this also persists at two loops \cite{Tamarit:2013nna})\footnote{For $q^2\gg\Lambda^2$, the EFT description is far more optimistic than the UV complete theory, which in part explains our decision to employ a method of truncation. Furthermore, the authors of Ref.~\cite{Primulando:2015lfa} found that, even with the large discrepancies in the spectra between the EFT and UV completion, the difference in coupling sensitivity of the different methods is $\mathcal{O}(25-50\%)$. Given the large uncertainties in other aspects of our proposed analysis, this is within the tolerance of our projections.}. Since our proposed analysis has excellent sensitivity even in this conservative limit, and the regions of EFT invalidity are often well-covered by other constraints from Sec.~\ref{sec:constraints}, we present our results in the EFT limit.  It is still an interesting problem to understand the reach of our search in a particular UV complete theory.

Additionally, in the limit where the EFT is not always valid at the scale of LHC parton-level interactions, direct searches for electrically charged states that induce $\mu_\gamma$ can become important \cite{Weiner:2012gm,Liu:2013gba}, but  charged states below 100 GeV are severely constrained by LEP. There may be overlap between such direct searches and the MiDM constraints for large dipoles, but the constraints from charged state searches are model-dependent.

\subsubsection*{$B$ factories}

As with the dark photon, the $B$-factories can have impressive coverage of the MiDM thermal relic parameter space for DM masses below a few GeV. 
If the photon from the $\chi_2$ decay registers in the detector, then signatures of MiDM can appear in prompt or displaced photon searches via $e^+e^-\rightarrow\gamma+\chi_2+\chi_1$. As before, we conservatively assume that events are recorded via the BaBar monophoton trigger.\\

\noindent{\bf Two prompt photons and missing energy:} For decay lengths $\ell_{\chi_2} < 1$ cm, the photon from the decay $\chi_2\rightarrow \chi_1+\gamma$ appears as a prompt photon. Thus in this case, the signal of interest is two prompt photons, one of which is used to pass the monophoton trigger, and missing energy arising from the decay of the neutral $\chi_2$. In particular, we consider a signal region with a leading photon satisfying $E_\gamma > 2~\GeV$ to pass the trigger, a sub-leading photon with $E>20~\MeV$ and missing energy $\slashed{E} > 50~\MeV$. This scenario inherits low backgrounds from the monophoton search, and we assume that such a signal region is background-free in our projections in Fig.~\ref{fig:mainplot-dipole}, as the dominant background is $\gamma\gamma$ and can be reduced with a missing momentum cut.  Thus, we use 10 signal events as a benchmark for the sensitivity to this model. Note that we apply the same acceptance cuts on the leading photon $-0.31 < \cos\theta_\gamma^* < 0.6$ (calculated in the CM frame) as Ref.~\cite{Aubert:2008as}. This centrality requirement was important for the analysis from Ref.~\cite{Aubert:2008as}, since the energy resolution is degraded at smaller $\theta_\gamma^*$, but less crucial for the MiDM signal as this final state no longer calls for a bump search. Relaxing this requirement could indeed result in a stronger sensitivity to this scenario.\\

\noindent{\bf One prompt photon, one non-pointing photon and missing energy:} An orthogonal final state to consider has a decay length  $\ell_{\chi_2} > 1$ cm. We conservatively only consider scenarios where the decay occurs within $\ell_{\chi_2} < 50$ cm. The signal region we define in this scenario is one prompt photon with $E_\gamma > 2~\GeV$, and one non-pointing photon with $E_\gamma > 20~\MeV$. As for the case above, we assume a background-free scenario in this signal region for our projection; we find the signal sensitivity to be sub-dominant to the diphoton + $\slashed{E}$ signal, although it may be useful in the case of unexpected backgrounds for the prompt search.

\section{Inelastic DM  Relic Abundances}
\label{sec:freeze-out}

In the standard thermal freeze-out scenario, the DM abundance today is fixed by the DM number density at the time that its annihilation rate into SM states falls below the Hubble expansion rate. For any given model, the observed relic abundance singles out a particular region in parameter space that is an obvious target for experimental searches. In many models, the thermal relic parameters are challenging to directly probe at colliders.

As we have seen, inelastic DM models offer additional handles at colliders to probe small couplings. To assess the sensitivity to the thermal relic cross section, we discuss in this section the calculation of the DM abundance in the (M)iDM models from Sec.~\ref{sec:models}, leading to the conclusion that thermal scenarios can be probed in many of the searches proposed in Sec.~\ref{sec:collider-searches}. In iDM models, DM annihilation proceeds predominantly via co-annihilation of the mass eigenstates; since $\chi_2$ is heavier, its abundance depletes faster than $\chi_1$ due to decays ($\chi_2 \to \chi_1 +\mathrm{SM}$) and scattering ($\chi_2 +\mathrm{SM}\to\chi_1+\mathrm{SM}$), and so freeze-out occurs earlier for mass splittings larger than the freeze-out temperature. This suggests that thermal relic iDM models feature larger couplings to compensate for the less efficient freeze-out.

 In terms of the variable $x\equiv m_2/T$, the  coupled Boltzmann equations for the DM densities $Y_i$ (normalized to the entropy density)  are:
\be\label{eq:main-bolts}
\frac{dY_{1,2}}{dx}  = - \frac{\lambda^{12}_{A}}{x^2} \biggl[  Y_1 Y_2 -     Y^{(0)}_1 Y^{(0)}_ 2\biggr] &&\nonumber \\ 
- \frac{\lambda_A^{11, 22} }{x^2 } \biggl[  Y_{1,2}^2-     (Y^0_{1,2})^2   \biggr] && \nonumber \\
  \pm \frac{\lambda_{S}}{x^2} Y^{(0)}_f   \biggl[    Y_2  -   \frac{Y_2^{(0)}}{Y_1^{(0)}} Y_1   \biggr] &&  
\nonumber \\ 
 \pm x   \gamma_{D}    \left[ Y_2  -     \frac{Y_2^{(0)}}{Y_1^{(0)}} Y_1   \right] &&  ~,~~
\ee
where $Y_i \equiv n_i /s$ is the comoving number density of each species, a $(0)$ superscript denotes an equilibrium quantity,   $s(T) = 2 \pi^2 g_{s,*}T^3/45$ is the entropy density, and $\lambda_A$, $\lambda_S$, and $\gamma_D$ are dimensionless annihilation, scattering, and decay rates respectively. $g_{s,*}(T)$ is the number of entropic degrees of freedom. The first line of the right-hand side characterizes the change in DM density due to co-annihilation, the second line gives the change due to self-annihilation, and the third and fourth lines characterize scattering and decay processes that keep $\chi_1$ and $\chi_2$ in chemical equilibrium with one another and in kinetic equilibrium with the SM.
Using the  Hubble rate during radiation domination $H(T) = 1.66 \sqrt{g_*} T^2/m_{P\ell}$ ($g_*$ is the number of relativistic degrees of freedom), the
dimensionless rates are defined to be 
\be
\lambda_A^{ij} &=&        \frac{s(m_2)}{H(m_2)}   \langle    \sigma v(\chi_i \chi_j \to SM) \rangle   \\
\lambda_{S} &=&  \frac{s(m_2)}{H(m_2)}       \langle \sigma v(\chi_2 f \to \chi_1f) \rangle    \\
\gamma_D &=&  H(m_2)^{-1}  \langle \Gamma( \chi_2 \to \chi_1 + SM ) \rangle, 
\ee
for $\chi_1\chi_2$ co-annihilation,  $\chi_2 f \to \chi_1 f$ 
 inelastic scattering, and $\chi_2 \to \chi_1 $ + SM decays respectively. 
The diagonal rate $\lambda_A^{ii}$ is non-zero if there exist processes that allow $\chi_i\chi_i\to\mathrm{SM}+\mathrm{SM}$ annihilation.

For the dark photon model, the scalar dark matter scenario is purely inelastic and so $\lambda_A^{ii}=0$. For fermion DM, 
there exists a self-annihilation channel whose rate is proportional to the 
difference of Majorana masses in Eq.~(\ref{eq:general-splitting}), and is also $p$-wave (helicity) suppressed for the SM vector (SM axial) current. For the pure dipole scenario,  the $\chi_i \chi_i \to \gamma \gamma$, $\gamma Z$, and $ZZ$ 
channels are always open if kinematically accessible, but the self-annihilation rate is suppressed by additional powers of the dipole moment.

As in most co-annihilation scenarios, the scattering/decay processes preserve kinetic and chemical equilibrium between $\chi_2$ and $\chi_1$ throughout freeze-out, and so the system of Boltzmann equations for $Y_{1,2}$ can be replaced by a single Boltzmann equation for $Y_{\rm tot} = Y_1+Y_2$,
\be
\frac{dY_{\rm tot}}{dx}= -\biggl[  \frac{Y_{\rm tot}^2}{(Y^{(0)}_{1}+Y_2^{(0)})^2 }-  1   \biggr]\sum_{i,j}\frac{\lambda^{ij}_A}{x^2}\,Y_i^{(0)}Y_j^{(0)}.
\ee
This approximation is valid over our parameter space.

\begin{figure}[t] 
 \hspace{-0.6cm}
\includegraphics[width=8.5cm]{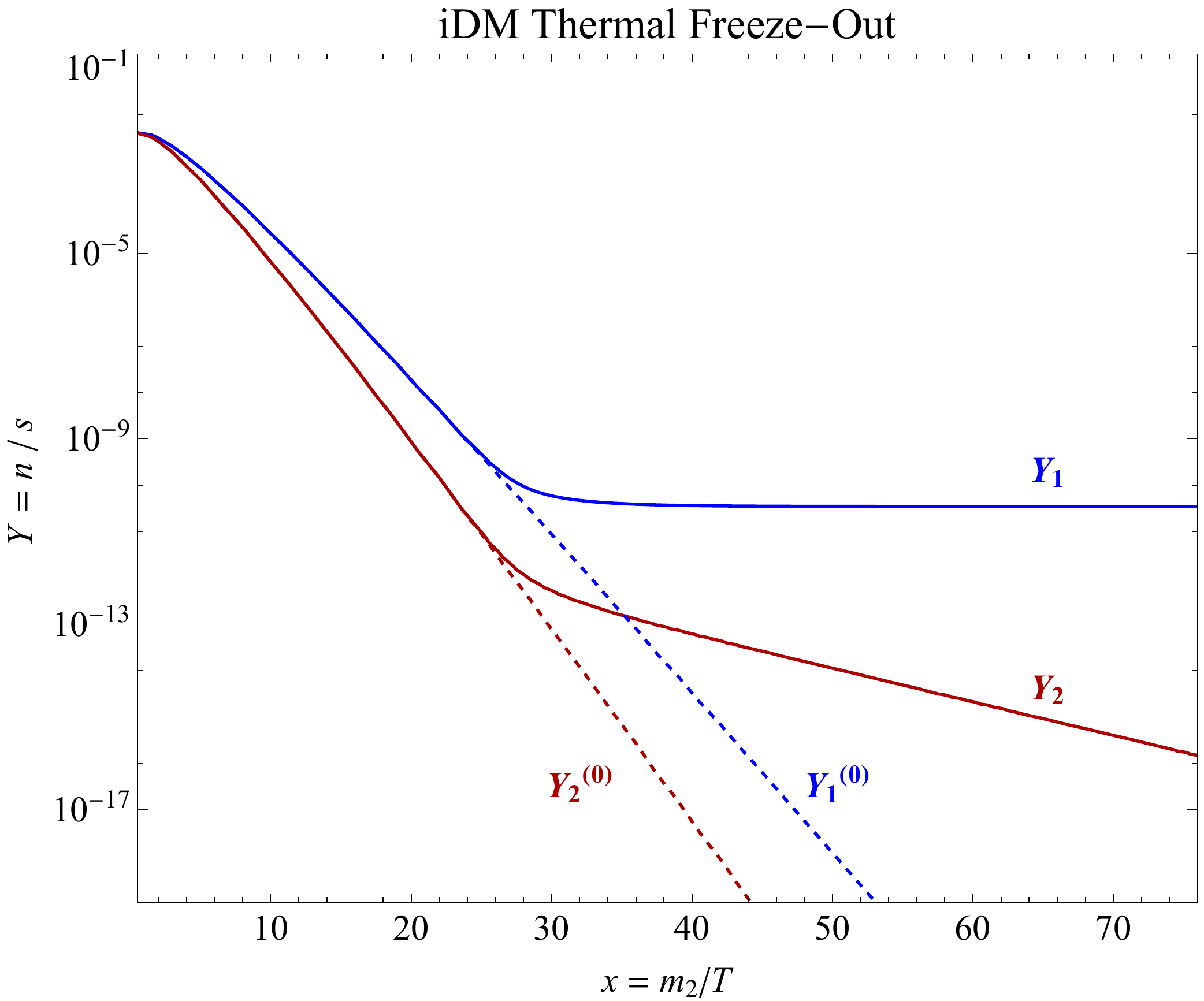}  
\caption{Freeze-out for fermion iDM (including co-annihilation and sub-dominant self-annihilation) mediated by an  $s$-channel $\apr$ with $m_1 = 10$ GeV, 
$\Delta = 0.2\, m_1$, 
and $m_{\apr} =  3\, m_1$ with $\langle \sigma v\rangle \sim 10^{-24} \cm^3 \,\mathrm{s}^{-1}$, for which
$\Omega_{\chi_1} \sim \Omega_{\rm DM}$ at late times. The solid (dashed) curves represent the actual (equilibrium) number densities for the $\chi_{1,2}$ species and we define 
the dimensionless evolution parameter $x \equiv m_2/T$.
 Note that  the excited state continues to steadily decay and down-scatter into $\chi_1$ off SM particles even after
 $\chi_1$ has frozen out.} \label{fig:freezeout}
\end{figure}

 Considering an example  point in the dark photon model, we show in Fig.~\ref{fig:freezeout} the $\chi_1$ and $\chi_2$ yields as a function of $m_2/T$. For each model, we determine the parameters of the theory that give the observed DM relic abundance as a function of $m_1$, and we show these curves in Figs.~\ref{fig:mainplot-fermion}-\ref{fig:mainplot-dipole}. We provide more comprehensive information on the rates that appear in the Boltzmann equations in Appendix \ref{sec:appendixB-annihilation-rate}.

\section{Constraints}\label{sec:constraints}

In this section, we consider  other constraints on the parameter space of the dark photon and dipole models, reviewing those which are complementary to  collider searches and those which are ineffective in iDM models. These probes include direct detection experiments, precision measurements of SM parameters, indirect detection, and LEP.

\subsection{  Precision Electroweak and QED Measurements }

For models with new neutral gauge interactions, mixing between the massive gauge bosons can lead to shifts in observed SM electroweak couplings that are excluded by electroweak precision and other observables.\\

\noindent {\bf Dark photon:} At high energies, the gauge invariant kinetic mixing term is between the $\apr$ and hypercharge. At low energies, this leads to mixing with the $Z$ and can shift the observed mass at the 
$Z$ pole to an unacceptable extent. Precision electroweak measurements are therefore sensitive to $\apr$ independent of its decay modes. Constraints from LEP \cite{Hook:2010tw} impose
a generic constraint on $\epsilon \lsim 10^{-2}$ at low $m_{\apr}$ and a stronger limit near the $Z$ mass. 
We show the exclusion region in Figs.~\ref{fig:mainplot-fermion} -- \ref{fig:mainplot-scalar}. 

There are additional constraints from the observed value of the muon's anomalous magnetic dipole moment, $(g-2)_\mu$. Despite
the $\sim 3\sigma$ discrepancy between theory and observation of $(g-2)_\mu$, the $5 \sigma$ bounds can still be used to set a conservative
limit on the $\epsilon/m_{\apr}$ ratio. In terms of $a_\mu =  (g-2)_\mu/2$, a kinetically mixed gauge boson induces a shift \cite{Pospelov:2008zw}
\be
\Delta a_\mu = \frac{\alpha \epsilon^2}{2\pi} \int_{0}^{1} dz \frac{2z (1-z)^2}{(1- z)^2 +  \left(  m_{\apr}/m_\mu  \right)^2 z}~,~
\ee
which constrains the upper-left portion of the parameter space in Figs~\ref{fig:mainplot-fermion} -- \ref{fig:mainplot-scalar}.\\

\noindent {\bf MiDM:} Since MiDM does not introduce any new gauge boson eigenstates, there are no relevant constraints on the coupling from precision SM measurements.

\subsection{LEP Constraints}

\noindent {\bf Dark photon:} iDM can be produced via $Z\rightarrow \chi_2\chi_1$, as well as via radiative return production of $\apr\rightarrow\chi_2\chi_1$. We find that the LEP searches for soft leptons and missing energy are subdominant to the $Z$-mass constraint discussed earlier.\\

\noindent {\bf MiDM:} We considered several constraints, including the $Z$ invisible width and direct searches for final states with photons. Over much of the parameter space, we find that the most powerful constraint comes from a LEP-1 search for two photons + missing energy \cite{Acton:1993kp}. There is excellent sensitivity to MiDM due to the relatively low photon thresholds (two central photons with $E_\gamma>1$) and the very low backgrounds (no events observed with $m_{\gamma\gamma}>5~\GeV$). We also include the constraint arising from the $Z\rightarrow \chi_2\chi_1$  contribution to the total width \cite{Agashe:2014kda}, which is particularly important for fractional mass splittings below 10\% where the diphoton + MET search is less effective\footnote{The photon from $\chi_2\rightarrow\chi_1+\gamma$ decay is often not soft enough to avoid the veto used in $Z$ invisible decay searches, and so $Z$ invisible width constraints  are typically subdominant to the total width measurement.} . 

The excluded region is shown in Fig.~\ref{fig:mainplot-dipole}, covering parts of the thermal relic line for larger splittings ($\Delta \gtrsim0.2\,m_1$). The LEP searches have also  been used to constrain models with dipole transitions between multiple Dirac DM states, giving rise to hard photons \cite{Primulando:2015lfa}. 

\subsection{Indirect Detection}

Null results in searches for DM annihilation in cosmic rays or at the epoch of recombination are typically very constraining of  elastic thermal DM models with $s$-wave annihilation to states other than neutrinos; for example, measurements of the cosmic microwave background (CMB) typically exclude such models for DM masses below $\sim10$ GeV \cite{Finkbeiner:2011dx,Galli:2011rz,Slatyer:2015jla}. In iDM models, however, the relic abundance is largely set by co-annihilation of DM states in the early universe, whereas the $\chi_2$ abundance is depleted today, suppressing indirect detection signals; the mass splittings we consider are too large for signals from collisional excitation \cite{Finkbeiner:2007kk}.  Nevertheless, sub-leading processes may mediate residual $\chi_1$ annihilation today, and we examine the constraints.\\

\noindent {\bf Dark photon:} In the scenarios we consider, $m_1 <m_{\apr}$, and in the limit of zero splitting, self-annihilation only occurs in either four-body final states or loop-induced processes, both of which lead to tiny indirect detection cross sections. If $\Delta$ is larger, self-annihilation can also occur through $p$-wave-suppressed (via mixing with $\gamma$ or $Z$) or helicity-suppressed (via mixing with $Z$) operators, both of which give suppressions that render indirect detection signals negligible. Therefore, there are no relevant constraints on this scenario.\\

\noindent {\bf MiDM:} Processes with two insertions of the dipole operator allow for the tree-level annihilation $\chi_1\chi_1\rightarrow\gamma\gamma$ and $\gamma Z$ (see Fig.~\ref{fig:DM_dip_ann}) \cite{Weiner:2012cb}. Dark matter annihilating  via this process gives rise to mono-energetic gamma rays originating from the Galactic Center. Even though the indirect detection cross section is suppressed relative to the single-photon process determining the relic abundance, gamma ray line searches have sensitivity to sub-thermal cross sections over a broad range of masses. We recast the limits from {\it Fermi's} line search, Ref.~\cite{Ackermann:2015lka}, and show the result in Fig.~\ref{fig:mainplot-dipole}. There is considerable uncertainty in the DM annihilation rate due to the unknown halo profile. To avoid over-stating the strength of the indirect detection bounds relative to collider searches, we use the Isothermal profile bounds in our recast, which is a factor of few weaker than an NFW profile.

Measurements of the CMB also place constraints on DM annihilation. DM annihilations near redshift $z \sim 10^3$  can inject energy into the visible sector and ionize hydrogen at the CMB's surface
of last scattering \cite{Finkbeiner:2011dx,Galli:2011rz,Slatyer:2015jla}. However, we find that  current constraints from photon line searches are stronger than those from the CMB, and so we do not include CMB constraints in our plots.

\subsection{LHC}
Searches during Run 1 at the LHC in principle can already set constraints on the representative models of iDM we consider. Here we discuss them in turn.\\

\noindent {\bf Dark photon:} We checked that standard monojet+$X$+$\slashed{E_T}$ searches for DM at the LHC don't constrain new parameter space for the dark photon iDM model. However, a recent CMS analysis aimed at stop pair production in SUSY which searched for low multiplicity of jets, in addition to a pair of soft prompt dileptons and missing energy, sets limits on the parameter space of our model \cite{CMS:2015eoa}. In particular we find it provides complementary coverage to the searches we propose, namely for larger $m_1$ and $\Delta$, as shown in Figs.~\ref{fig:mainplot-fermion} -- \ref{fig:mainplot-scalar}.\\ 

\noindent {\bf MiDM:}  We have verified that potentially constraining searches such as a diphoton and missing energy analysis by ATLAS from Ref.~\cite{ATLAS:2014uca} do not cover new parameter space in the MiDM model.\\

\subsection{ Direct Detection }

One of the principal motivations for iDM models is that $\chi_1$-nucleus tree-level elastic scattering $\chi_1 N \rightarrow \chi_1 N$ is no longer the dominant scattering reaction;  the corresponding inelastic process, $\chi_1N\rightarrow \chi_2 N$ is only allowed for certain kinematic configurations, modifying the na\"ive predictions of DM spin-independent scattering \cite{TuckerSmith:2001hy}. In particular, when $\Delta \gg |\vec{q}\,|^2 / 2M_N$ for momentum transfer $|\vec{q}|$ and target mass $M_N$, the inelastic process is kinematically forbidden due to the low DM velocity. Tree-level spin-independent scattering no longer occurs, significantly reducing all direct detection constraints.


\begin{figure}[t!]
 \vspace{0.cm}
\includegraphics[width=5.2cm]{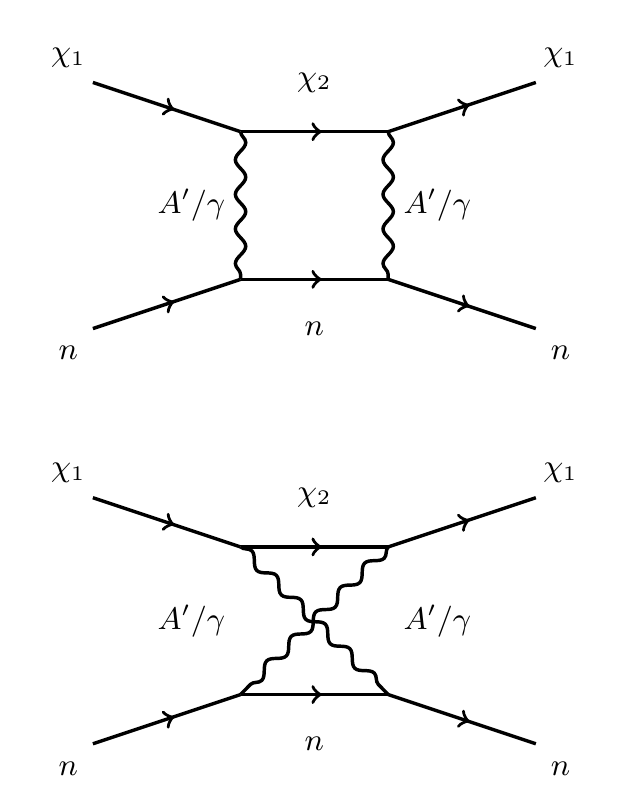}  ~
\caption{ Leading elastic scattering diagram for  DM that couples off-diagonally to a vector mediator or to a photon via 
the dipole interaction in the MiDM scenario. For $\Delta$ much larger than the kinetic energy, we can contract the $\chi_2$ internal
line and use the effective interaction shown in Fig.~\ref{fig:dipscat}.}
\label{fig:fermion_dd_box}
\end{figure}


With spin-independent $\chi_1$-nucleon cross section limits now in the vicinity of $\sim10^{-45}\,\,\mathrm{cm}^2$ \cite{Aprile:2012nq,Akerib:2013tjd}, other processes contributing to elastic DM scattering can also lead to constraints. These include higher-order loop-induced processes, as well as on-diagonal interactions suppressed by the mass splitting. Here, we give an overview of the various constraints, with details of the calculations provided in  Appendix \ref{sec:appendix-direct-detection}.\\

\begin{figure}[t!]
 \vspace{0.cm}
\includegraphics[width=5.2cm]{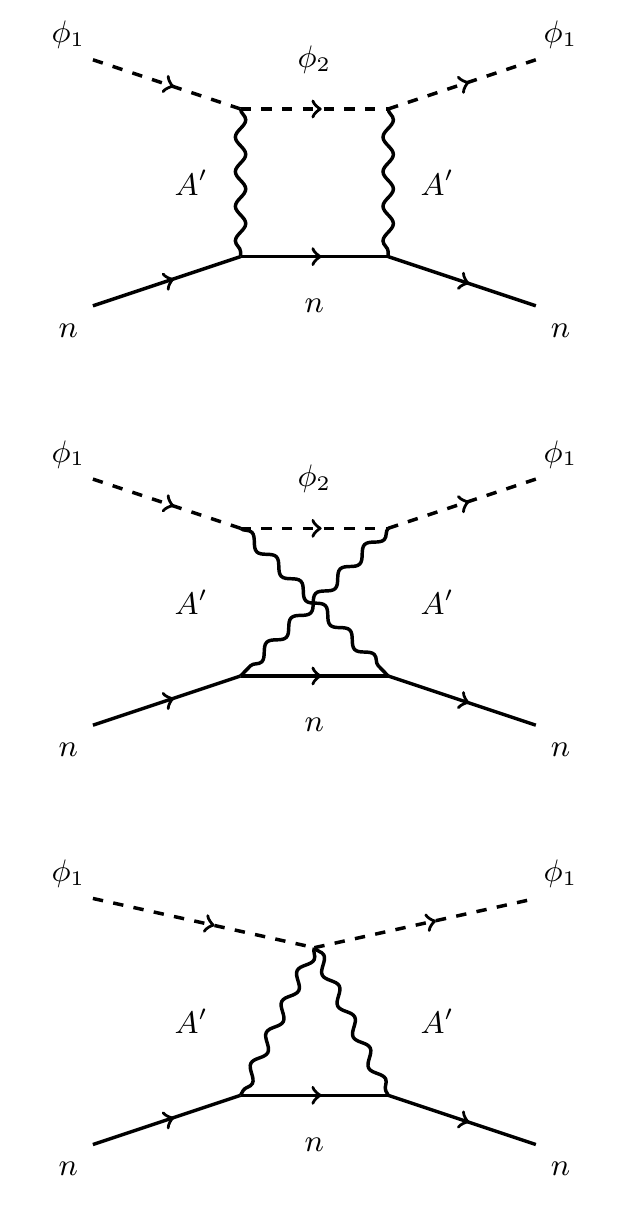}  ~
\caption{ Leading elastic scattering diagram for scalar DM that couples off-diagonally to a vector mediator. Note that the $\phi_1\phi_1\apr \apr$ vertex
induces a third diagram at the same order. }
\label{fig:scalarbox}
\end{figure}

\noindent {\bf Dark photon:} Elastic $\chi_1$-nucleus scattering can occur at loop level by exchanging pairs of dark photons via a virtual $\chi_2$ state (see Figs.~\ref{fig:fermion_dd_box} and \ref{fig:scalarbox}). Whether elastic scattering enjoys a coherent nuclear enhancement depends on the characteristic energy scale of the dark photons:~the contribution peaks from loop momenta $\sim m_{\apr}$, and given a nuclear radius $R_N$, a coherent enhancement occurs for $m_{\apr} \ll R_N^{-1}$ and is suppressed for $m_{\apr} \gg R_N^{-1}$  \cite{Batell:2009vb}. In this latter case, the DM resolves nuclear substructure and scatters predominantly off nucleons. In this paper, we consider $m_{\apr} > m_\chi\gtrsim100$ MeV, and so we are nearly always in the regime where scattering occurs off of nucleons. Consequently, the elastic scattering rate is not only loop suppressed, but also suppressed by $Z^2$ relative to conventional spin-independent scattering. For the parameter space of our study, we find that loop-induced direct detection is never important relative to other constraints on the dark photon model for splittings $\Delta/m_1\sim0.1$, although direct detection could be more relevant for smaller splittings where collider constraints are less effective.

In models where the Majorana masses for each DM charge eigenstate are maximally different ($|m_\eta-m_\xi|=m_\eta+m_\xi$ in the notation of Sec.~\ref{sec:models}), on-diagonal couplings to the dark photon are present as shown in  Eq.~(\ref{eq:general-splitting}). This gives rise to elastic scattering at leading order in $\epsilon$, but suppressed by the mass splitting $\Delta$ and the DM squared velocity. For $m_{\apr}$ masses well below the $Z$, such that the dark photon mixes predominantly with the photon, the spin-independent cross section per-nucleon at zero momentum transfer in the limit of small splitting is
  \be
  \sigma_{\chi n} \simeq \frac{16 \pi \epsilon^2 \alpha \alpha_D \Delta^2Z^2}{ m_1^2 m_{\apr}^4A^2} \mu_{\chi n}^2 v^2~,~~
  \ee
  where $v$ is the DM velocity and  $Z (A)$ is the atomic (mass) number of the target. We find that, due to the velocity suppression, the limits from LUX \cite{Akerib:2013tjd} are subdominant to other constraints. For $m_{\apr}\gtrsim m_Z$, an additional spin-dependent, velocity un-suppressed channel opens due to coupling with the SM axial current, but this constraint is also sub-dominant.\\

\begin{figure}[t!]
 \vspace{0.5cm}
  \hspace{0.5cm}
\includegraphics[width=6cm]{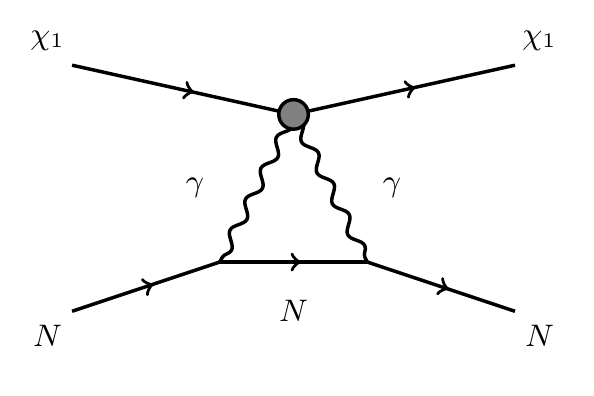}  ~
\caption{ Leading elastic scattering diagram for direct detection in the MiDM scenario with the $\chi_2$ excited state integrated out. The 
gray circle represents an insertion of the effective  operator from Eq.~(\ref{eq:rayleigh}). }
\label{fig:dipscat}
\end{figure}

\noindent {\bf MiDM:} The dipole interaction  induces elastic $\chi_1$-nucleus scattering at loop level (see Figs.~\ref{fig:fermion_dd_box} and \ref{fig:dipscat}), and scattering can occur off of both nuclear charge and dipole \cite{Chang:2010en,Banks:2010eh}. In the limit where the splitting is sufficiently large that $\chi_2$ cannot be produced on-shell,  the leading effective interaction can be written in terms of a Rayleigh operator \cite{Weiner:2012cb}
\be\label{eq:rayleigh}
{\cal L} \simeq \biggl( \frac{\mu_\gamma^2 }{2 m_2}  \biggr) \, \overline \chi_1 \chi_1 F_{\mu \nu} F^{\mu \nu}~,~~
\ee
where we have integrated out the excited state $\chi_2$ (we neglect additional velocity-suppressed terms). The direct detection scattering rate for this operator was computed in Ref.~\cite{Weiner:2012cb}.  The resulting spin-independent, elastic
cross section {\it per nucleon} is
\be
\sigma_n^{\rm (SI)}  \approx \frac{  \alpha^2 Z^4    \mu_{\chi n}^2}{ \pi^2 A^2} \frac{   \mu^4_\gamma  }{ m_2^2  } \, Q_0^2~,~~
\ee   
where  $Z (A)$ is the atomic (mass) number of the target, $Q_0 \sim \sqrt{6}(0.3+0.89A^{1/3}) \,\,{\rm fm}^{-1}$ is the nuclear
coherence scale, and $\mu_{\chi n}$ is the DM-nucleon reduced mass. See Appendix \ref{sec:appendix-direct-detection} for a full discussion. We find that the LUX \cite{Akerib:2013tjd} constraint is subdominant over the parameter space.


\section{Conclusions and Discussion}\label{sec:conclusions}

In this article, we have proposed a set of searches at colliders that seek to exploit the distinctive signatures of inelastic dark matter (iDM). In iDM scenarios, direct and indirect detection constraints are weaker than for elastic DM, and therefore one prominent  blind spot of current efforts is in the DM mass range of $\sim 100~\MeV-100~\GeV$. For fractional mass splittings between the two DM states of order $\sim 10\%$, the heavier excited DM state can decay into the lighter state along with SM particles at decay lengths typically smaller than the size of collider experiments. These states can be reconstructed at colliders, giving rise to distinctive iDM signatures.

These additional handles of iDM at colliders allow for targeted searches that may allow for sensitivity down to couplings favoured by a thermal DM origin in some of the simplest iDM scenarios. In particular, we focused on two representative models, described in Sec.~\ref{sec:models}:~that of a split DM system accompanied by a dark photon that kinetically mixes with the SM, and that of magnetic dipole iDM. For each model, we propose new, powerful collider searches that can greatly increase sensitivity:~in the dark photon model, we find that searches for monojet + missing momentum + displaced leptons could cover nearly all of the remaining thermal relic territory for DM masses between $\sim 100~\MeV-100~\GeV$ and splittings of $\mathcal{O}(10\%)$. For MiDM, the proposed searches for monojet + missing momentum + photon can also cover unexplored parameter space compatible with thermal freeze-out (for comparable masses and  splittings as the dark photon model).

In both scenarios, the thermal relic parameter space for much larger fractional splittings is essentially excluded by other experiments. For smaller splittings and DM masses below a few GeV, we find that monophoton + missing momentum searches at Belle II could provide complementary sensitivity in this regime \cite{Essig:2013vha}. Finally,  we comment that significant strong sensitivity could be achieved for $m_{\rm DM} \lesssim \rm{few}~\GeV$ and small splittings with  new  electron beam-dump and active-target experiments \cite{Izaguirre:2013uxa,Izaguirre:2014dua, Izaguirre:2014bca,Izaguirre:2015yja,Izaguirre:2015pva}, proposed proton-beam fixed target experiments \cite{deNiverville:2011it,Dharmapalan:2012xp,Kahn:2014sra},  and future direct detection experiments \cite{Cushman:2013zza} optimized for low threshold sensitivity.

{\em Acknowledgments.}
We thank Chris Brust, Christopher S. Hill, Bingxuan Liu, Maxim Pospelov, Philip Schuster, Abner Soffer, Tim Tait, Natalia Toro, Wells Wulsin, and Itay Yavin for helpful conversations. 
The Perimeter Institute for Theoretical Physics is supported by the Government of Canada through Industry Canada
and by the Province of Ontario. EI receives partial support from the Ministry of Research and Innovation - ERA (Early Research Awards) program.

\appendix

\section{Freeze-out Rates}
\label{sec:appendixB-annihilation-rate}

In this Appendix, we summarize the relevant cross sections and rates for the thermal relic abundances computed in Sec.~\ref{sec:freeze-out}. We provide these rates to illustrate their characteristic scaling; consequently, we typically show only leading terms in an expansion in powers of DM velocity and mass splitting. For simplicity, we also present dark photon rates in the limit $m_\apr \ll m_Z$, in which case we can assume that the dark photon couples predominantly to the electromagnetic current (this limit is also valid over much of the parameter space we consider).

\subsection{Annihilation Rates}
For the $\apr$ mediated interactions, the leading $s$-wave annihilation rate for fermion iDM is
\be
\sigma v(\chi_1 \chi_2 \! \to  \bar ff ) &\approx& 
     \frac{      4 \pi \epsilon^2 \alpha \alpha_D   (m_1+m_2)^2    }{   [(m_1+m_2)^2-m_{A'}^2]^2} ,
\ee
where we have omitted corrections of order $m_f/m_1$. 
For scalar iDM, the analogous $\apr$ mediated process is $p$-wave suppressed 
\be
\sigma v(\phi_1 \phi_2 \to \bar f f) &=& 
     \frac{    8 \pi \epsilon^2 \alpha \alpha_D  m_1m_2 }{  3 [(m_1+m_2)^2-m^2_{A'}]^2}\,v^2 ,
\ee
where $v$ is the relative velocity.

For the dipole mediated scenario, including both $\gamma$ and $Z$ dipole operators, the $s$-channel annihilation rate into SM fermions $\bar f f$
correct to first order in splitting and velocity 
is  \cite{Weiner:2012cb}
 \be \label{eq:dip-annihilation}
 \hspace{-0.cm}\sigma v (\chi_1 \chi_2 \to  f\bar f) &=& \alpha Q_f^2  \mu_\gamma^2 \biggl(  1 + 2 v_f  \frac{\mu_Z}{\mu_\gamma}\zeta[(m_1+m_2)^2]  
 \nonumber \\ && + (v_f^2    +    a_f^2) \frac{\mu^2_Z}{\mu^2_\gamma} \zeta^2[(m_1+m_2)^2]  \biggr)~,~~~~~~
 \ee
where $Q_f$ is the fermion electromagnetic charge, $v (a_f)$ is the ratio of  fermion's vector (axial) charges to the 
electromagnetic charge, and  $\zeta(s) \equiv s/(s -m_Z^2).$

Over the DM mass window we consider in this paper ($100~\MeV -100$ GeV), annihilation to hadronic resonances 
can play a key role in setting the relic abundance in the early universe, yielding the 
spikes in the relic density curves in Figs.~\ref{fig:mainplot-fermion}-\ref{fig:mainplot-dipole}. We can approximately account for the analytically intractable
 phase space for hadronic final states by using the known ratio of hadron and muon production
in $e^+e^-$ annihilation $R(s) \equiv \sigma(e^+e^- \to {\rm hadrons})/\sigma(e^+e^- \to \mu^+\mu^-)$
from Ref.~\cite{Agashe:2014kda}. We decompose the full DM annihilation rate into SM states according to 
\be
&& \hspace{-0cm} \sigma v_{\rm full}  =  (\sigma v)_{e^+e^-} \theta(s -  4m_e^2)  \nonumber \\ && ~~ + (\sigma v)_{\mu^+\mu^-}  \biggl[  
  \theta(s -  4m_\mu^2)  +  \theta(s -  4m_\pi^2) R(s)  \biggr]~.~~~~~~
\ee
This is done in the vicinity of the hadronic resonances, $\sqrt{s}\lesssim12~\GeV$. For larger DM masses, the couplings to partons are used. 

The $R(s)$ approximation is  valid only for  couplings to the SM vector current; the SM axial  current has a contribution to the cross section which differs from the vector current by terms proportional to $m_f^2$, where $f$ is the final state fermion.  In the dark photon model, the axial coupling only arises from mixing with the $Z$ and scales like $m_{\apr}^2/m_Z^2$ for $m_{\apr}\ll m_Z$; since the $R(s)$ approximation is only used for $m_\chi\lesssim$ 6 GeV, and we consider $m_{\apr}$ not parameterically larger than $m_\chi$, the axial coupling is negligible and the $R(s)$ approximation is valid. Similarly, for the dipole the $Z$ contribution is suppressed by $s^2/m_Z^4$ in the $R(s)$ regime, and so is negligible.

\subsection{Scattering Rates}
For both fermion and scalar DM, scattering through a virtual $\apr$ mediator
has the same parametric dependence 
\be
\langle \sigma v(\chi_2 e \to \chi_1e) \rangle &\sim& 16 \pi \epsilon^2  \alpha \alpha_D \frac{T^2 }{   m_{\apr}^4 }.~~~
\ee
For fermion DM scattering through the dipole interaction, the leading order term in the velocity expansion is 
\be
\langle \sigma v(\chi_2 e \to \chi_1e) \rangle &\sim& 3\alpha \mu_\gamma^2.~~~~~
\ee 
In the above, the relative velocity $v \sim 1$ because  the electrons remain relativistic. For typical values of the model parameters, these cross sections lead to rates larger than the Hubble expansion.

\subsection{Decay Rates}
For fermion DM in the $\apr$ mediated scenario, the $\chi_2$ width for $\Delta \ll m_1,\,m_{\apr}$  is 
\be
 \label{eq:fermion-width}
\Gamma(\chi_2 \to \chi_1 ~e^+e^-)  = \frac{4 \epsilon^2 \alpha \alpha_D \Delta^5 }{15 \pi m_{\apr}^4 }.  ~~~
\ee
The corresponding expression for scalar DM de-exciting through a virtual $\apr$  has the same value in the $\Delta \ll m_{\apr}$ limit.

 For excited dark-fermions decaying in the MiDM model, the  width is 
\be
\Gamma(\chi_2 \to \chi_1  \gamma) = \frac{ \mu_\gamma^2 \Delta^3}{\pi}~.~~
\ee 

In each scenario, the width is thermally averaged in the conventional way to obtain $\langle\gamma_2\rangle$.

\subsection{Inelastic Thermal Averaging}
To perform the  thermal average for iDM in each scenario, we generalize the results of 
Ref.~\cite{Gondolo:1990dk} to obtain to first order in the splitting $\Delta$:
\be
&& \langle \sigma v \rangle = \frac{1}{N(T)} \int_{s_0}^\infty ds \, \sigma  (s - s_0)\sqrt{s}  K_1 \left( \frac{\sqrt{s}}{T}\right) ~,~~~ 
\ee
where $s_0=(m_1+m_2)^2$ and the averaging factor is 
\be
N(T)  = 8 m_1^2 m_2^2 \, T K_2\left(  \frac{m_1}{T} \right) K_2\left(  \frac{m_2}{T} \right) ~.~  
\ee
$K_1$ and $K_2$ are the modified Bessel functions of the first and second kind.

\section{Direct Detection}
\label{sec:appendix-direct-detection}

\subsection{$\apr$ Fermion iDM Direct Detection }

At loop level, iDM mediated by a massive $\apr$ can scatter at direct detection experiments via the box diagram
in Fig.~\ref{fig:fermion_dd_box}. For $m_{\apr} \gg R_N^{-1}$, the 1-loop Feynman diagram in Fig.~\ref{fig:fermion_dd_box} involves a single nucleon on the SM side (as opposed to a nucleus). The differential cross section can be defined as
\be
\frac{d\sigma}{d\Omega} =  \frac{\mu_{\chi n}^2}{4\pi^2} |I(m_\chi, \Delta)|^2,
\ee
where $I$ is the integral over the loop momentum and  $\mu_{\chi n}$ is the $\chi$-nucleon reduced mass. This can be evaluated using
the formalism of Heavy Quark Effective Theory (HQET) as in Ref.~\cite{Weiner:2012cb}.
 For zero momentum
transfer, the integral  can be written as
\be
I(m_\chi, \Delta) \simeq  \frac{8 i  \epsilon^2 \alpha \alpha_D }{ \Delta }    
\int_0^\infty d| \vec \ell \, | |\vec \ell \, |^2 
\frac{F( |\vec \ell| )^2}{ ( |  \ell |^2 + m_{\apr}^2 )^{2}}; ~~ 
\ee
for simplicity, we show the result only for the $\apr$ coupling to the electromagnetic current, which is valid in the $m_\apr \ll m_Z$ limit. 

In contrast with Refs.~\cite{Batell:2009vb,Weiner:2012cb},  the $\apr$ in our regime of interest generates a short range force that 
resolves nuclear substructure, so the scattering is predominantly off nucleons without the additional $Z^4$ enhancement
from coherent, loop-scattering off nuclear targets. The monopole form-factor for scattering off
protons can be parameterized by
$ F(q) = (1+   q^2/ m_p^2  )^{-2} $, and analogous form factors can be defined as well for neutrons  \cite{deNiverville:2011it,AguilarArevalo:2010cx}. For the masses and splittings
considered in this paper, existing direct detection constraints from LUX are 
too weak to constrain this process.

\subsection{$\apr$ Scalar iDM Direct Detection }
The scattering amplitude for this process is the sum of diagrams in Fig.~\ref{fig:scalarbox}. To leading
order, the two box diagrams can be written in the HQET limit  as \cite{Weiner:2012cb}
\be
&& {\cal A}_{1} + {\cal A}_{2}\simeq
\frac{ g_D^2 e^2 \epsilon^2 (\overline u u)_n}{ \Delta m_1}  \int \frac{d^3 | \vec \ell \, | }{(2\pi)^3}  F(|\vec \ell\,|)^2 \int_0^\infty d{\cal E} \nonumber\\ && \hspace{1cm} \times  \int_{-\infty}^\infty \frac{d\ell^0}{2\pi} \frac{    (\ell^0)^2 +  (2 m_1-   {\cal E}/2)^2}{[ (\ell^0)^2 - \lambda({\cal E})]^3},
\ee
where $\cal E$ is a dimensionful Feynman parameter introduced to simplify denominator products of unequal mass dimension,
$F(q) = (1 + q^2/m_p)^{-2}$ is the proton form factor introduced above, and $\lambda(\mathcal{E}) = \mathcal{E}^2/4 + m_{\apr}^2+|\vec{\ell}\,|^2$.

The third (triangle) diagram contributes
\be
{\cal A}_3 
\simeq g_D^2 e^2 \epsilon^2 ( \overline u u )_n   \int_0^\infty \! \! \! \!  d{\cal E} \int  \! \frac{d^3 |\vec \ell |}{(2\pi)^3} \int_{-\infty}^\infty \!  \frac{d\ell^0}{2\pi} \frac{F(|\vec \ell\,|)^2}{      [ (\ell^0)^2 - \lambda({\cal E})]^3      } ,~~~ \nonumber
\ee
so the total amplitude ${\cal A} = \sum_i {\cal A}_i$ is squared to give 
\be
\langle |{\cal A}|^2 \rangle = 
4 m_n^2 |I(m_1 ,\Delta)|^2,
\ee
where the integral is 
\be 
&& \hspace{-0.5 cm}I(m_1,\Delta) = 
-  \frac{8 i \alpha \alpha_D   \epsilon^2}{\Delta m_1}  
 \int_0^\infty \!\!\!  d| \vec \ell \, | |\vec \ell \, |^2 
  \Biggl\{         \biggl(     \frac{4m_1^2}{ (|\vec \ell\,|^2 + m_{\apr}^2)^2} 
\nonumber  \\ &&  \hspace{-0.7cm} - \frac{2 m_1}{ (|\vec \ell\,|^2 + m_{\apr}^2)^{3/2}} + \frac{1}{ |\vec \ell\,|^2 + m_{\apr}^2} 
   \biggr)   +  
  \frac{   \Delta m_1 }{4 (|\vec \ell\, |^2 + m_{\apr}^2)^2}  \Biggr\}   F(|\vec \ell\,|)^2 \nonumber \\ 
\ee
and we have averaged over the spin of the nucleon target. The $\phi_1$-nucleon cross section is now
\be
 \sigma_{\phi n} = \frac{|I(m_1,\Delta) |^2m_n^2}{ 4 \pi (m_1 + m_n)^2}.
\ee

\subsection{Magnetic Dipole Direct Detection}

For magnetic dipole scattering through the same boxes in Fig.~\ref{fig:dipscat}, we can work in the limit where $ \Delta$ is much
larger than the kinetic energy of the DM, and we integrate
out the excited state. This is equivalent to performing an HQET expansion on the DM propagator. The leading-order effective interactions between DM and nuclei arises from two photon interactions, which at leading order in the expansion and up to Lorentz index contractions are
\be\label{eq:raydm}
{\cal L} = \biggl(      \frac{\mu_\gamma^2 }{2 m_2}  \biggr) \left(   \, \overline \chi_1 \chi_1 F_{\mu \nu} F^{\mu \nu}  + 
i\overline \chi_1 \gamma^5 \chi_1 F_{\mu \nu} \tilde{F}^{\mu \nu}    \right) ~,~~
\ee
where $\tilde{F}_{\mu \nu} \equiv (1/2) \epsilon_{\mu\nu\alpha\beta} F^{\alpha \beta}$ is the dual electromagnetic field-strength tensor. 
To match the conventions in \cite{Weiner:2012cb} for this limit, we identify $g_{\gamma\gamma}/4 \Lambda^3_R
 \equiv \mu_\gamma^2/2m_2$. At one loop level, these interaction give rise to elastic scattering at one loop level
 through the diagram depicted in Fig.~\ref{fig:dipscat}; however, the contribution through scattering from the second operator in Eq.~(\ref{eq:raydm}) is
 velocity suppressed and can be neglected when computing direct detection rates.  
 
 The differential cross section for coherent non-relativistic scattering off nuclear target $N$ is 
\be
\frac{d\sigma}{d\cos\theta} = \frac{\mu_{\chi N}^2 }{2 \pi} \left|   \frac{\alpha Z^2   \mu_\gamma^2 Q_0}{ 2m_2    }  F_N\left(\frac{| \vec q \, |^2}{Q_0^2}\right) \right|^2 ,~~
\ee
where $\vec q$ is the momentum transfer,  $\mu_{\chi N} \equiv m_1 m_N / (m_1 + m_N)$ is the DM-nucleus reduced mass, and 
$Q_0 \equiv \sqrt{6} (0.3 + A^{1/3})^{-1} {\rm fm}^{-1}$ is the coherence scale for a target of mass number $A$. In this limit the spin-independent 
 cross-section {\it per nucleon} can be approximated as 
\be
\sigma_n^{\rm (SI)}  \simeq \frac{  \alpha^2 Z^4    \mu_{\chi n}^2}{ \pi^2 A^2} \frac{   \mu^4_\gamma  }{ m_2^2  } \, Q_0^2~.~~
\ee   
where $\mu_{\chi n}$ DM-{\it nucleon} reduced mass and 
and we have approximated the Helm form factor \cite{Helm:1956zz} as $F_N(0) \simeq 2 /\sqrt \pi$ at zero momentum transfer. Note that this
form is only valid in the limit where the mass splitting is much larger than the DM kinetic energy, which is trivially satisfied in our regime of interest throughout this paper.

\bibliographystyle{apsrevM}
\bibliography{LHC-iDM}

\end{document}